\documentclass[seceq]{ptptex}
\usepackage{graphicx}

\title{
Systematic solution-generation of five-dimensional black holes%
}

\author{
Hideo \textsc{Iguchi},$^{1,}$\footnote{e-mail:
iguchi.h@phys.ge.cst.nihon-u.ac.jp}
Keisuke \textsc{Izumi}$^{2,}$\footnote{e-mail:
izumi@yukawa.kyoto-u.ac.jp}
and
Takashi \textsc{Mishima}$^{1,}$\footnote{e-mail:
tmishima@phys.ge.cst.nihon-u.ac.jp}
}

\inst{$^1$ Laboratory of Physics,~College of Science and Technology,~
Nihon University,\\ Narashinodai,~Funabashi,~Chiba 274-8501,~Japan
\\
$^2$ Yukawa Institute for Theoretical Physics, 
Kyoto University, Kyoto 606-8502, Japan}

\abst{%
Solitonic solution-generating methods are powerful tools to construct nontrivial black hole solutions of the higher-dimensional Einstein equations systematically. 
In five dimensions particularly, the solitonic methods can be successfully applied to the construction of asymptotically Minkowski spacetimes with multiple horizons. 
We review the solitonic methods applicable to higher-dimensional vacuum spacetimes and present some five-dimensional examples derived from the methods. 
 
}

\begin{document}

\maketitle
\tableofcontents
\section{Introduction}

\subsection{Outline}
In the legendary era of general relativity from the 1970s to the early 1980s the solution-generating techniques have been greatly developed in four dimensions and applied to construct new series of axisymmetric stationary solutions extensively: See Ref.~\citen{Stephani:2003tm} for a comprehensive review. 
Particularly the practical and useful solitonic methods based on the integrability of axisymmetric and stationary gravitational field equations were derived by several authors. 
For example, the B\"{a}cklund transformation analogues were invented by Harrison~\cite{ref2} and Neugebauer~\cite{ref3}, 
and also a certain kind of inverse scattering method was developed through the successive works by Belinsky and Zakharov  \cite{Belinsky:1971nt}. 

In short, in both the methods a known solution (i.e., a seed) is transformed into a new one by adding `solitons' which have formally a role analogous to ordinary solitons appeared in general nonlinear physics. 
In these methods the solitons can be added in principle with arbitrary number so that a sequence of an infinite number of solutions is generated systematically. 
In axisymmetric and stationary cases, nontrivial solitons can bring rotations into the gravitational systems.  
As one of remarkable applications of these methods, Kramer and Neugebauer succeeded to generate so called double-Kerr solutions~\cite{Kramer:1980} as vacuum solutions of Einstein equations, and opened a new way to the study of gravitational systems with multiple horizons since the work done by Israel and Khan based on the Weyl solutions~\cite{Isreal:1964}.
In four dimensions the multi-horizon vacuum systems in asymptotically flat spacetimes generally have some flaws like conical singularities.  
So introducing spin-spin repulsion effects by adding nontrivial solitons has been expected to remove such flaws.   
It seems, however, that such completely regular multiple-black-hole spacetimes can not be realized in four dimensions~\cite{Neugebauer:2009}.   

Mainly in the 1980s, several pioneering works of more than four dimensional 
spacetimes have been done.
They have also introduced the solitonic methods to analyze the highe-dimensional gravity in the context of the standard Kaluza-Klein theory with compactified extra dimensions,  
and further developed studies of higher-dimensional gravitational objects like K-K black holes to some extent~\cite{refKK}. 
However, the pace of advancing the study was rather slow because of the increasing intricacy of the methods in higher dimensions and also the absence of new stimulus after the rise of the unified theories in particle physics including the first revolutionary period of string theory. 

The recent discovery of the solutions of five-dimensional black ring, especially rotating regular black ring~\cite{{Emparan:2001wn}} has injected new life into all the related subjects of higher-dimensional gravity. 
Inspired by this, systematic construction of new solutions of five-dimensional black holes has also progressed remarkably using solitonic methods. 

The recent progress has been done mainly along two ways, though mutual interactions between them often occurred. 
First the solitonic method based on the Neugebauer's B\"{a}cklund transformation~\cite{Neugebauer:1980} was applied to the five-dimensional Mikowski spacetime as an appropriate seed, and the solutions of black ring with rotation in $S^2$ were discovered~\cite{Mishima:2005id,Iguchi:2006tu}.
This method was successively applied to several seeds, and generated several interesting solutions ($S^1$-rotating black ring~\cite{Iguchi:2006rd} and multiple black hole systems like black di-ring~\cite{Iguchi:2007is}). 
However, its application is rather restricted due to the metric form adopted to make the method available. 
Actually, as described in section 3.1 the method can only deal with the spacetimes whose metrics have only one off-diagonal component. 
That is, the black holes with two independent rotations cannot be constructed by this method. 
Soon after the above attempt the original Belinsky and Zakharov's inverse scattering method was also used to generate some of the above solutions~\cite{Tomizawa:2005wv,Tomizawa:2006vp}. 
However, the original inverse scattering method also suffered from the same restriction as that of the former method (see \S 2.1). 

Following the above works an ingenious device was introduced by Pomeransky~\cite{Pomeransky:2005sj} to make the inverse scattering method fully effective in higher dimensions ( see \S 2.2 for details of this idea). 
In this work five-dimensional Myers and Perry black holes with two rotations were reconstructed. 
Once the powerful method given, several interesting solutions have been generated with the inverse scattering method. 
The multiple black hole systems with single rotation, so called black Saturn and black di-ring, have been also constructed by this method~\cite{Elvang:2007rd,Evslin:2007fv}, though these solutions can be derived using the former methods because the metric sector corresponding to the three Killing vectors has the structure of 2+1 block matrix. 
Moreover, the solution of a double Myers-Perry black hole is constructed by
the inverse scattering method~\cite{Herdeiro:2008en}.
The inverse scattering method shows its power when the off-diagonal components generally appear in the metric sector corresponding to the Killing vectors.  
That is, when the doubly spinning rings or rotational black lenses must be considered.  
The solutions of doubly spinning rings and rotational black lenses were 
constructed in Ref.~\citen{Pomeransky:2006bd} and Ref.~\citen{Chen}, respectively. 
As another nontrivial example, the orthogonal black di-rings (bicycling black rings) were also constructed and the regularities and other physical properties were investigated~\cite{Izumi:2007qx}. 

The solution-generating tasks will continue, because there remain a lot of interesting solutions to be dealt with by using the solitonic methods. 
So at present it may be useful to present a summary of solitonic methods and exhibit some of the generated solutions. 
In the following, for short of space, we concentrate on solitonic solution-generating methods corresponding to higher-dimensional vacuum Einstein equations and the axisymmetric and stationary spacetimes that asymptotically behave like a Minkowski space.
We just comment here that there are several works where they have tried to generate the solutions of other systems like Einstein-Maxwell~\cite{Ida:2003,Yazadjiev:2008}, 
Einstein-Maxwell-Chern-Simon~\cite{Figueras:2010} and 
Einstein-Maxwell-Dilaton~\cite{Chng:2008sr} systems.
After a brief introduction of the rod structure below, \S\S 2 and 3 are devoted to describing two different solitonic methods mainly used so far: the inverse scattering method and the B\"{a}cklund transformation analogue, respectively. 
In \S\S 4 and 5 the procedures to generate solutions with single horizon and disconnected horizons are demonstrated, respectively. Here, only the procedures of the inverse scattering method are shown. 

\vskip 0.3cm
\noindent
{\it Notations and Conventions}
\vskip 0.2cm
Through this chapter we adopt a higher-dimensional canonical metric form to describe the $D$-dimensional stationary spacetimes with $D-2$ commutable Killing vectors (see the metric form (\ref{4-metric}) in \S 2.1).
The coordinates associated with the $D-2$ Killing vectors are denoted by $x^{i}$. 
Two-dimensional subspace orthogonal to the Killing vectors is described with $\rho$ and $z$.
The ranges of $\rho$ and $z$ are $0 < \rho$ and $-\infty < z < \infty$ respectively.
The Latin indices $i$, $j$ and $k$ which run from 1 to $D-2$ are used to assign the Killing vectors to the quantities ($g_{ij}$, $\Psi_{ij}$, $\mu_{i}$ ...). 
The Latin indices $p$ and $q$ are kept for labeling the quantities with solitons. 
Both the indices $1$ and $t$ mean time components.
Especially in five dimensions, $\phi$ and $\psi$ are used in place of $x^{2}$ and $x^{3}$. 
And also in \S 3.1, 
the prolate-spheroidal coordinates $x$ and $y$ are introduced, which are defined by 
$
\,\rho=\sigma \sqrt{x^2-1} \sqrt{1-y^2},\ z=\sigma xy\,
$
with the ranges $1\le x$ and $-1 \le y \le 1$.
%

\subsection{Rod structure}
%
The rod structure was first deliberately used to study the so called generalized Weyl solutions of D-dimensional vacuum Einstein equations, which admit $D-2$ orthogonal commuting Killing vector fields~\cite{Emparan:2001wk}, and next  extended to the stationary and axisymmetric cases~\cite{Harmark:2004rm}.
In recent progress of solitonic solution-generations, the rod structure and practically the diagrammatic view of the rod structure become powerful tools.    
In fact, as examples shown in the following sections, using the analysis of rod structures we can guess and construct appropriate seed solutions to generate a desired solution.  
Also mathematical and physical properties of the resultant solution can be investigated from the viewpoint of rod structure with the aid of mathematical facts like the uniqueness theorems of solutions~\cite{Hollands:2008,Armas:2010}. 
%

Originally `rods' appeared in the process to construct the Weyl or generalized Weyl solutions as the rod sources of  gravitational potential in hypothetical three-dimensional Newtonian gravity. 
For the static and axisymmetric case with the $D-2$ orthogonal Killing vector fields, the metric admits the diagonal form 
\begin{equation}
ds^2=-e^{2U_1(\rho,z)}(dx^{1})^2+\sum_{i=2}^{D-2} e^{2U_{i}(\rho,z)}(dx^{i})^2+e^{2\nu(\rho,z)}(d\rho^{2} +dz^2).
\end{equation}
The functions $U_{i}$ are axisymmetric solutions of the hypothetical three-dimensional Laplace equations
\begin{equation}
\left( \partial_{\rho}^{2}+\frac{1}{\rho}\partial_{\rho}+\partial_{z}^2 \right)U_{i}=0,
\label{1-a}
\end{equation}
for $i=1,\cdots ,D-2$, and also satisfy the constraint derived from the Einstein equations 
\begin{equation}
\sum_{i=1}^{D-2}U_{i}=\ln\rho.
\label{1-b}
\end{equation}
Once a set of $\left\{U_{1},\cdots,U_{D-2}\right\}$ given, the quantity $\nu$ can be determined by integrating the following equations
\begin{eqnarray}
\partial_{\rho}\nu
 &=&-\frac{1}{2\rho}+\frac{\rho}{2}\sum_{i=1}^{D-2}\left[(\partial_{\rho}U_{i})^{2}-(\partial_{z}U_{i})^{2}\right], \\
\partial_{z}\nu
 &=&\rho\sum_{i=1}^{D-2}\partial_{\rho}U_{i}\partial_{z}U_{i}\,.
\end{eqnarray}
From Eqs. (\ref{1-a}) and (\ref{1-b}), the functions $U_{i}$ can be considered as the Newtonian potentials produced by thin rods lying at some positions on $z$-axis, and all $U_{i}$ must add up to the Newtonian potential corresponding to an infinite thin rod with linear mass density $1/2$. 
We can say that when a set of rods satisfies the constraint (\ref{1-b}) and each rod is assigned to one of Killing vector fields, a corresponding solution can be constructed uniquely. 
In this place, it should be, however, noticed that each rod does not necessarily have linear mass density $1/2$, so that regularity of the corresponding generalized Weyl solution cannot be guaranteed. 
To construct a regular solution in which occurrence of conical singularities is allowed, each rod must also have linear mass density $1/2$. 
When all of the rods have $1/2$ density, equation (\ref{1-b}) means that there is only one rod corresponding to every points on the $z$-axis except some isolated points (i.e., junctions of adjacent rods). 
On each rod in this case, the metric coefficient $e^{U_{i}}$ associated with the Killing vector field which is assigned to the rod behaves $O(\rho^2)$, so that the interval on the $z$-axis corresponding to the rod becomes a rotational axis or a horizon in the corresponding spacetime. 

We now introduce the rod structure of the solution as follows:
first prepare the set of rods that was made by dividing $z$-axis into appropriate intervals; 
next assign a $(D-2)$-dimensional vector (so called the direction vector) to each rod:
\begin{equation}
v=v^{i}\frac{\partial}{\partial x^{i}}.
\end{equation}
The direction vector $v$ in the generalized Weyl solutions just corresponds to one of Killing vectors 
$
\frac{\partial}{\partial x^{i}}.
$
On the interval corresponding to the rod, the norm of $v$ is zero:
\begin{equation}
|v|^2=g_{ij}(0,z)v^{i}v^{j}=0\,.
\end{equation}
Extension of rod structure to stationary and axisymmetric solutions with off-diagonal metric components can be done almost similarly: see Ref.~\citen{Harmark:2004rm} for complete discussion. 
In general case the direction vector $v=v^{i}\frac{\partial}{\partial x^{i}}$ can be defined through 
\begin{equation}
g_{ij}(0,z)v^{j}=0\,.
\end{equation}
There being off-diagonal metric components, the direction becomes mixed with some Killing vectors. 
For example, the direction vector for the rod associated with a horizon is $(1, \Omega_{2}, \cdots, \Omega_{D-2})$ where $\Omega_{i}$ is a angular velocity of $i$-th rotation of the horizon, and for the rod associated with normal rotational axis the direction is $(0, \cdots, 1, \cdots, 0)$ where 1 appears in $i$-th slot ($i=2,\cdots,D-2$). 
Furthermore as an interesting but more complicated case, the rod structure of the spacetime with black lenses was also considered~\cite{Evslin:2008}. 

\section{Solution-generating technique I: Inverse scattering method}

\subsection{Procedure of inverse scattering method}

In this section, we show the technique of constructing vacuum solutions, 
inverse scattering method~\cite{Belinsky:1971nt,Tomizawa:2005wv}.
This technique can be applied only if
we have $D-2$ commutable Killing vectors in $D$-dimensional spacetime. 
In the inverse scattering method, a known solution 
is transformed into a new solution. 
Here, we concentrate on how to construct vacuum solutions of spacetime and 
do not prove that vacuum Einstein equations are satisfied 
in the obtained new solutions.
The detailed proof is shown in the original paper~\cite{Belinsky:1971nt}.

We begin with the $D$-dimensional stationary spacetime with $D-2$ commutable 
Killing vectors $(\partial/\partial x^1)$, $(\partial/\partial x^2)$, $\cdots$, 
$(\partial/\partial x^{D-2})$, where $(\partial/\partial x^1)$ is a timelike Killing 
vector field and $(\partial/\partial x^2)$, $\cdots$, $(\partial/\partial x^{D-2})$ are 
spacelike Killing vector fields. 
According to Harmark's paper~\cite{Harmark:2004rm}, 
without loss of generality the metric can be written as 
\begin{eqnarray}
ds^2=f(\rho,z)(d\rho^2 +dz^2)+g_{ij}(\rho,z)dx^i dx^j,
\label{4-metric}
\end{eqnarray}
where $f(\rho,z)$ and $g_{ij}(\rho,z)$ are a scalar function and 
an induced metric on $(D-2)$-dimensional hypersurface, respectively, 
and the induced metric satisfies $\det g_{ij}=-\rho^2$. 
Einstein equations of this metric become
\begin{eqnarray}
&&\partial_\rho U_{i}^{\ j} + \partial_z V_{i}^{\ j}=0,
\label{4-geq} \\
&& U_{i}^{\ j}= \rho(\partial_\rho g_{ik} ) g^{kj}, 
\label{4-Udef}\\
&&V_{i}^{\ j}= \rho(\partial_z g_{ik} ) g^{kj}, 
\label{4-Vdef}\\
&&\partial_\rho \ln f = - \frac{1}{\rho} +\frac{1}{4\rho}
(U_{i}^{\ j} U_{j}^{\ i} - V_{i}^{\ j} V_{j}^{\ i}), 
\label{4-rhof}\\
&&\partial_z \ln f =  \frac{1}{2\rho} U_{i}^{\ j} V_{j}^{\ i}.
\label{4-zf} 
\end{eqnarray}
The first three equations (\ref{4-geq})-(\ref{4-Vdef}) 
are a set of non-linear differential equations relating to only 
the induced metric $g_{ij}$. 
In the last two equations (\ref{4-rhof}) and (\ref{4-zf}), 
derivatives are operated on only the scalar function $\ln f$.
In order to obtain solutions, we solve the first three non-linear equations for the 
induced metric $g_{ij}$, and then, integrate the last two equations.
While the integration of the last two equations is not difficult, 
it is hard to solve the first three equations because of non-linearity. 
Applying the technique of solving the inverse problem in the scattering physics, 
we can solve the first three equations. 

In the inverse scattering method, we must prepare a seed metric which 
is an already known solution of vacuum Einstein equations. 
Here, we suppose that we have a seed metric $g^{(0)}_{ij}$.
The way to construct a seed metric is shown in the next subsection. 
Defining a generating matrix $\Psi^{(0)}_{ij}(\lambda,\rho,z)$,
we construct the linear differential equations in $\Psi^{(0)}_{ij}(\lambda,\rho,z)$ as
\begin{eqnarray}
&&\left( \partial_z - \frac{2\lambda^2}{\lambda^2 + \rho^2}\partial_\lambda
\right) \Psi^{(0)}_{ij} =
\frac{\rho V_{\ \ i}^{(0) k}-\lambda U_{\ \ i }^{(0) k}}{\lambda^2+\rho^2}
\Psi_{kj}^{(0)} ,
\label{4-Psi1}\\
&&\left( \partial_\rho + \frac{2\lambda \rho}{\lambda^2 + \rho^2}\partial_\lambda
\right) \Psi^{(0)}_{ij} =
\frac{\rho U_{\ \ i}^{(0) k}-\lambda V_{\ \ i }^{(0) k}}{\lambda^2+\rho^2}
\Psi_{kj}^{(0)},
\label{4-Psi}
\end{eqnarray}
where $U_{\ \ i}^{(0) k}$ and $V_{\ \ i}^{(0) k}$ are 
$U_{i}^{\ k}$ and $V_{i}^{\ k}$ made of $g^{(0)}_{ij}$, respectively, and 
$\lambda$ is a new complex parameter independent of $\rho$ and $z$.

First, we must solve these equations for $\Psi^{(0)}_{ij}$, 
since the new solutions are described with $\Psi^{(0)}_{ij}$ 
and any parameters. 
Equations (\ref{4-Psi1}) and (\ref{4-Psi}) are easier to solve than 
Einstein Eqs. (\ref{4-geq})-(\ref{4-Vdef}) 
because they are linear equations in $\Psi^{(0)}_{ij}$. 

Secondly, we introduce functions 
\begin{eqnarray}
&&\mu_p(\rho,z) = \sqrt{\rho^2+(z-a_p)}-(z-a_p),\\
&&\bar \mu_q(\rho,z) = -\sqrt{\rho^2+(z-a_q)}-(z-a_q),
\end{eqnarray}
where $a_p$ is a real constant. 
$\mu_p(\rho,z)$ and $\bar \mu_q(\rho,z)$ are called a soliton and an anti-soliton, 
respectively. 
In this section, 
we represent solitons $\mu_p$ and anti-solitons $\bar \mu_p$ as ${\mu'}_p$, collectively. 
We also introduce $(D-2)$-dimensional vectors $m^{(p)}$ associated with ${\mu'}_p$.
The $(D-2)$-dimensional vectors $m^{(p)}$ are called BZ vectors. 

Next, we construct $n\times n$ matrix as 
\begin{eqnarray}
\Gamma_{pq}=
\frac{m^{(p)}_i\left(  \Psi^{(0)}({\mu'}_p,\rho,z)^{-1} \right)^{ij}
g^{(0)}_{jk} \left( \Psi^{(0)}({\mu'}_q,\rho,z)^{-1} \right)^{kl}
m^{(q)}_l}
{\rho^2+{\mu'}_p{\mu'}_q}, 
\end{eqnarray}
where $n$ is the number of solitons and anti-solitons we introduce.

Finally, a new metric is described as
\begin{eqnarray}
&&g_{ij}= \rho^{-2n/(D-2)} 
\left(\prod_{p=1}^n {\mu'}_p^{2/(D-2)}\right){g'}_{ij},
\label{4-newg}\\
&&{g'}_{ij}=\left(
g^{(0)}_{ij}- \sum_{p,q}(\Gamma^{-1})^{pq} {\mu'}_p^{-1}{\mu'}_q^{-1}
N_i^{(p)} N_j^{(q)}
\right), 
\label{4-newg'}\\
&&N_i^{(p)}= m_j^{(p)} \left( \Psi^{(0)}({\mu'}_p,\rho,z)^{-1}
\right) ^{jk}g^{(0)}_{ki}.
\label{4-Ni}
\end{eqnarray}
The obtained new metric $g_{ij}$ satisfies Eqs. (\ref{4-geq})-(\ref{4-Vdef}) 
and $\det {g_{ij}}=-\rho^2$. 
Actually, Eq. (\ref{4-newg'}) also satisfies Eqs. (\ref{4-geq})-(\ref{4-Vdef}). 
The factors $\rho^{-2n/(D-2)} \prod_{p=1}^n {\mu'}_p^{2/(D-2)}$ in Eq. (\ref{4-newg}) 
are needed to satisfy $\det {g_{ij}}=-\rho^2$.

A scalar function $f$ can be also constructed with the scalar function of 
the seed metric $f^{(0)}$. 
Integrating Eqs. (\ref{4-rhof}) and (\ref{4-zf}) with  Eq. (\ref{4-newg}), 
we know the new scalar function $f$ is 
\begin{eqnarray}
f= C f^{(0)} \rho^{\frac{-n^2}{D-2}}
\left( \prod_{p=1}^n {\mu'}_p^{\frac{2(n+D-3)}{D-2}} \right)
\left( \prod_{p,q=1,p>q}^n ({\mu'}_p-{\mu'}_q)^{\frac{-4}{D-2}} \right)
\det \Gamma_{pq} .
\label{4-newf}
\end{eqnarray}

The inverse scattering method can not be applied at the points where $\rho=0$. 
Equations (\ref{4-geq})-(\ref{4-Vdef}) are singular on $\rho=0$ because of 
$\det g_{ij}=-\rho^2$.
The singularities are coordinate singularities or physical singularities. 
Thus, there is no guarantee that the new solution is regular on $\rho=0$.
After the transformation, the structures at points where $\rho=0$ must be checked. 
If we can find that there are no physical singularities at points where $\rho=0$ 
(or if we can remove all physical singularities by tuning the parameters), 
we can expected that the obtained metric is a regular solution of the vacuum Einstein equations. 
In the opposite point of view, we don't care about singularities 
at points where $\rho=0$ in a seed metric. 
In order to construct a regular metric, 
we pay attention to the structures only in the new metric.

\subsection{Diagonal seed metric}

In the inverse scattering method, we must prepare a seed metric. 
In this section, we show a manner of constructing seed metrics. 
Here, we restrict our attention to a diagonal seed metric because it can be easily 
constructed.  

We start from the following diagonal metric
\begin{eqnarray}
g^{(0)}_{ij}= \mbox{diag}\left(
-g_1,g_2,\cdots,g_{D-2}
\right).
\end{eqnarray}
Then, each component is decoupled in Eqs. (\ref{4-geq})-(\ref{4-Vdef}) 
and we can solve them easily. 
The equations for the components are 
\begin{eqnarray}
\partial_\rho (\rho \partial_\rho \ln g_i) +
\partial_z (\rho \partial_z \ln g_i)=0 \qquad 
(i=1,\cdots,D-2) . \label{4-diag}
\end{eqnarray}
These equations can be solved because they are linear equations in $\ln g_i$. 
In fact, products of $\rho^n$, solitons and anti-solitons, 
that is,  
\begin{eqnarray}
g_i= \pm \rho^n \prod_{p=1}^{m} {\mu'}_p, 
\end{eqnarray}
where $n$ and $m$ are arbitrary numbers, 
satisfy Eq. (\ref{4-diag}) and hereinafter we use this solution. 
Because of $\det g_{ij}= -\rho^2$, diagonal metrics which are 
solutions of Einstein equations 
(except for on $\rho=0$) are represented as 
\begin{eqnarray}
g^{(0)}_{ij}= \mbox{diag}\left(
-\frac{{\mu'}_1{\mu'}_2 \cdots}{{\mu'}_3\cdots}, 
\frac{\rho^2 {\mu'}_4 \cdots}{{\mu'}_1\cdots},
\frac{{\mu'}_3\cdots}{{\mu'}_2{\mu'}_4\cdots},
\cdots
\right).
\label{4-seedg}
\end{eqnarray}
In sum, the way to construct diagonal metrics is to put minus sign ($-$) at the first component, 
$\rho^2$ at a numerator of a component, and solitons (and/or anti-solitons) 
at numerators and denominators so that the number of the same soliton (anti-soliton) 
at numerators is equal to that at denominators. 

The scalar function of the seed metric $f^{(0)}$ must be constructed. 
For ease in explanation of the method of constructing $f^{(0)}$, 
we use the following notation 
\begin{eqnarray}
g^{(0)}_{ij}= \mbox{diag}\left(
-\frac{{\mu'}_{1,1}^{(n)}{\mu'}_{1,2}^{(n)} \cdots}
{{\mu'}_{ 1,  1}^{(d)}{\mu'}_{1,2}^{(d)}\cdots}, 
\frac{{\mu'}_{2,1}^{(n)}{\mu'}_{2,2}^{(n)} \cdots}
{{\mu'}_{2,  1}^{(d)}{\mu'}_{2, 2}^{(d)}\cdots},
\cdots,
\frac{\rho^2{\mu'}_{k,1}^{(n)}{\mu'}_{k,2}^{(n)} \cdots}
{{\mu'}_{k,  1}^{(d)}{\mu'}_{k, 2}^{(d)}\cdots},
\cdots
\right),
\label{4-ex}
\end{eqnarray}
where, of course, this metric is constructed in the way just described 
in Eq. (\ref{4-seedg}) and the $k$-th component has the factor $\rho^2$.
Integrating Eqs. (\ref{4-rhof}) and (\ref{4-zf}) with Eq. (\ref{4-ex}), 
we can obtain the concrete form of $f^{(0)}$,
\begin{eqnarray}
&&f^{(0)}=k^2 \left(\prod_p \frac{{\mu'}_{k,p}^{(n)}}{{\mu'}_{k, p}^{(d)}}\right)
\nonumber\\
&& \qquad\times
\prod_{i=1}^{D-2}\Biggl\{
\left( \prod _{p, q} \frac{\rho^2+{\mu'}_{i,p}^{(n)}{\mu'}_{i,  q}^{(d)}}
{{\mu'}_{i,p}^{(n)}{\mu'}_{i,q}^{(d)}}  \right)
\left( \prod _{p\neq q} \frac{{\mu'}_{i,p}^{(n)}{\mu'}_{i,q}^{(n)}} 
{\rho^2+{\mu'}_{i,p}^{(n)}{\mu'}_{i,  q}^{(n)}} \right)
\left( \prod _{p\neq q} \frac{{\mu'}_{i,p}^{(d)}{\mu'}_{i,q}^{(d)}} 
{\rho^2+{\mu'}_{i,p}^{(d)}{\mu'}_{i,  q}^{(d)}} \right)
\nonumber\\ 
&&\qquad\qquad\qquad\qquad\qquad\times
\left( \prod _{p} \frac{\left({\mu'}_{i,p}^{(n)}\right)^2}
{\rho^2+\left({\mu'}_{i,p}^{(n)}\right)^2} \right) ^{\frac{1}{2}}
\left( \prod _{p} \frac{\left({\mu'}_{i,p}^{(d)}\right)^2}
{\rho^2+\left({\mu'}_{i,p}^{(d)}\right)^2} \right) ^{\frac{1}{2}}
\Biggr\} ,
\end{eqnarray}
where $k$ is an integration constant.

\subsection{Inverse scattering method with diagonal seed metric}

Starting from a diagonal seed metric and performing a double transformation 
make the calculation of obtaining the new metric easier. 
Advantage of starting a diagonal seed metric 
is easiness to solve Eq. (\ref{4-Psi}) for $\Psi_{ij}$. 
Advantage of performing the double transformation 
is that we don't need to care about the normalization factor 
in Eq. (\ref{4-newg}).
We have an additional advantage that the seed metric is easily predicted. 
Due to these advantages, we often start from a diagonal seed metric 
and perform the double transformation.

At first, we explain the double transformation. 
We denote the initial seed metric, the obtained metric 
after the first transformation 
and the final obtained metric as $g^{(0)}_{ij}$, ${\tilde g}^{(0)}_{ij}$ and $g_{ij}$, 
respectively. 
In the double transformation we don't need to care about the normalization factor 
in Eq. (\ref{4-newg}).
We will explain the reason later.
We prepare some solitons $\mu_p$ and anti-solitons $\bar \mu_q$ 
and make anti-solitons $\bar\mu_p$ and solitons $\mu_q$ corresponding to 
the solitons $\mu_p$ and the anti-solitons $\bar \mu_q$, respectively.
In the first transformation, starting from a diagonal seed metric
we use anti-solitons $\bar\mu_p$ and solitons $\mu_q$
with the trivial BZ vectors  which are equal to $(\partial/\partial t)$.
The soliton transformation (without attentiveness to the determinant) with 
trivial BZ vectors is 
the same as an operation of a multiplication only in the $tt$ component by 
$\prod_{p} (-\rho^2/{\bar\mu}_p^2)\prod_{q} (-\rho^2/{\mu}_q^2)$. 
Thus, ${\tilde g}^{(0)}_{ij}$ is written as 
\begin{eqnarray} 
{\tilde g}^{(0)}_{ij} = 
\mbox{diag}\left( \prod_{p}\left(\frac{-\rho^2}{{\bar\mu}_p^2}\right)\prod_{q} 
\left(\frac{-\rho^2}{{\mu}_q^2}\right) g^{(0)}_{tt}, 
g^{(0)}_{22},g^{(0)}_{33}, \cdots \right).
\end{eqnarray}
In the opposite point of view, the initial seed metric $g^{(0)}_{ij}$ can be obtained 
by multiplying ${\tilde g}^{(0)}_{ij}$ in only the $tt$ component by
\begin{eqnarray} 
\prod_{p} (-{\bar\mu}_p^2/\rho^2)\prod_{q} (-{\mu}_q^2/\rho^2)= 
\prod_{p} (-\rho^2/ {\mu}_p^2)\prod_{q} (-\rho^2/{\bar\mu}_q^2).
\end{eqnarray}
This means that the initial seed metric $g^{(0)}_{ij}$ can be obtained by 
the inverse scattering method with the solitons $\mu_p$ and the 
anti-solitons $\bar\mu_q$ starting from the metric ${\tilde g}^{(0)}_{ij}$. 
Therefore, we call the first transformation 
``removing the solitons $\mu_p$ and the anti-solitons $\bar \mu_q$". 
After that, we perform the second transformation. 
In the second transformation, starting from ${\tilde g}^{(0)}_{ij}$, we 
use the solitons $\mu_p$, anti-solitons ${\bar \mu}_q$ and non-trivial BZ vectors.

\begin{figure}[t]
  \begin{center}
    \includegraphics[keepaspectratio=true,height=65mm]{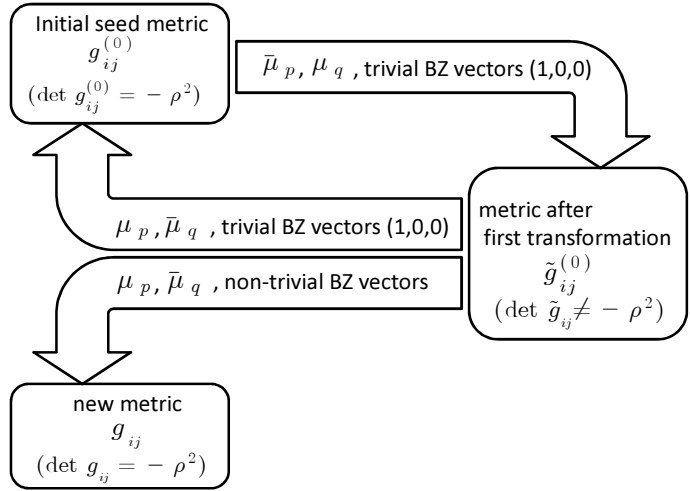}
  \end{center}
  \caption{
Double transformation; Starting an initial seed metric $g^{(0)}_{ij}$, 
we create the metric ${\tilde g}^{(0)}_{ij}$ by the inverse scattering method 
with anti-solitons $\bar \mu_p$, solitons $\mu_q$ and the trivial BZ vectors. 
Using a seed metric as the metric obtained in the operation of the 
first transformation ${\tilde g}^{(0)}_{ij}$, we perform the second 
transformation with solitons $\mu_p$, anti-solitons ${\tilde \mu}_q$ and BZ vectors.
Then, we obtain the new metric. 
If we use the trivial BZ vectors in the second transformation, 
the finally obtained metric is reverted to the initial seed metric.
}
  \label{4-fig1.eps}
\end{figure}

One of the advantages by starting from a diagonal metric is easiness
to solve the linear differential 
equations (\ref{4-Psi1}) and (\ref{4-Psi}).
In the inverse scattering method, we must solve the linear differential 
equations (\ref{4-Psi1}) and (\ref{4-Psi})
for $\Psi^{(0)}_{ij}$. 
Because of the diagonalization of a seed metric, 
Eqs. (\ref{4-Psi1}) and (\ref{4-Psi}) become diagonalized and can be solved easily. 
Complying the following guides, we can construct $\Psi^{(0)}_{ij}$.  
We represent the seed metric $g^{(0)}_{ij}$%
\footnote{
Since the first transformation is equivalent to the operation of the multiplication 
only in $tt$ component, 
we don't need to create $\Psi^{(0)}_{ij}$. 
In the second transformation,  
the seed metric is the metric after the first transformation ${\tilde g}^{(0)}_{ij}$, 
and thus, Eqs. (\ref{4-Psi1}) and (\ref{4-Psi}) is made of ${\tilde g}^{(0)}_{ij}$.
We denote ${\tilde \Psi}^{(0)}_{ij}$ as 
the solution of Eqs. (\ref{4-Psi1}) and (\ref{4-Psi}) in the second transformation 
when we concretely construct the solutions in \S \S \ref{4-Sec:single} 
and \ref{4-Sec:multi}. 
}
only with solitons and/or 
anti-solitons using the identity $\rho^2=- \mu_p {\bar \mu_p}$ and  
replace all solitons and anti-solitons ${\mu'}_p$ by $({\mu'}_p-\lambda)$.
The obtained matrix satisfies Eq. (\ref{4-Psi}), id est it is $\Psi^{(0)}_{ij}$. 
For example, when we deal with the seed metric (\ref{4-ex}), 
$\Psi^{(0)}_{ij}$ becomes
\begin{eqnarray}
&&\Psi^{(0)}_{ij}=\mbox{diag}\Biggl(
-\frac{({\mu'}_{1,1}^{(n)}-\lambda)({\mu'}_{1,2}^{(n)}-\lambda) \cdots}
{({\mu'}_{ 1,  1}^{(d)}-\lambda)({\mu'}_{1,2}^{(d)}-\lambda)\cdots}, 
\frac{({\mu'}_{2,1}^{(n)}-\lambda)({\mu'}_{2,2}^{(n)} -\lambda)\cdots}
{({\mu'}_{2,  1}^{(d)}-\lambda)({\mu'}_{2, 2}^{(d)}-\lambda)\cdots},
\cdots,
\nonumber\\
&&\qquad\qquad\qquad\qquad\quad
\frac{-({\mu}_{k,0}-\lambda)({\bar\mu}_{k,0}-\lambda)
({\mu'}_{k,1}^{(n)}-\lambda)({\mu'}_{k,2}^{(n)}-\lambda) \cdots}
{({\mu'}_{k,  1}^{(d)}-\lambda)({\mu'}_{k, 2}^{(d)}-\lambda)\cdots},
\cdots
\Biggr).
\end{eqnarray}

Performing the double transformation removes the burden of
normalization in  Eq. (\ref{4-newg}). 
The factors $\rho^{-2n/(D-2)} \left(\prod_{p=1}^n {\mu'}_p^{2/(D-2)}\right)$ 
in Eq. (\ref{4-newg}) are needed for tuning the determinant of a new metric $g_{ij}$ to $-\rho^2$.
This adjustment does not depend on BZ vectors, and thus,
the BZ vectors do not affect the determinant of the metric. 
 From the viewpoint of starting from the metric ${\tilde g}^{(0)}_{ij}$, 
the initial seed metric $g^{(0)}_{ij}$ and the new metric $g_{ij}$ are 
created by the transformation with the same solitons and anti-solitons 
while the BZ vectors are different. 
Thus, if we construct the initial seed metric $g^{(0)}_{ij}$ 
whose determinant is $-\rho^2$, 
that of the new metric $g_{ij}$ automatically becomes $-\rho^2$. 
In the same reason, regarding the construction of the scalar function of the new metric 
$f$ we do not need to mind the factors in Eq. (\ref{4-newf}) 
and the new scalar function $f$ is 
\begin{eqnarray}
f=Cf^{(0)} \frac{\det \Gamma_{pq}}{\det \Gamma^{(0)}_{pq}},
\label{4-f-new}
\end{eqnarray}
where $\Gamma_{pq}$ is created with ${\tilde g}^{(0)}_{ij}$, 
the solitons $\mu_p$, anti-solitons ${\bar \mu}_q$, 
and non-trivial BZ vectors, 
and $\Gamma^{(0)}_{pq}$ is created with ${\tilde g}^{(0)}_{ij}$, 
the solitons $\mu_p$, anti-solitons ${\bar \mu}_q$, 
and trivial BZ vectors.

The initial seed metric $g^{(0)}_{ij}$ belongs to the family of 
the new metric $g_{ij}$ obtained by the double transformation.
While the transformation with trivial BZ vectors by starting from the metric 
${\tilde g}^{(0)}_{ij}$ gives 
the initial seed metric $g^{(0)}_{ij}$, 
we can obtain the new metric $g_{ij}$ by the transformation with non-trivial 
BZ vectors. 
Parameters of the transformation are introduced via the BZ vectors and,
in this sense, the new metric $g_{ij}$ becomes the initial seed metric 
$g^{(0)}_{ij}$ by tuning these parameters. 
Therefore, it is expected that there are some similarities between the 
seed metric $g^{(0)}_{ij}$ and new metric $g_{ij}$.
Actually, in the constructions of the solutions in \S \S \ref{4-Sec:single} and 
\ref{4-Sec:multi}, 
we start from the seed metric whose rod structure is similar to 
that of the new metric (but unfortunately not the same).

\subsection{Construction of metric corresponding to given rod structure}

Hereinafter, we concentrate on solutions in five-dimensional spacetime. 
Moreover, since the inverse scattering method can be applied only to the 
case with three commutable Killing vectors, 
we suppose that we have them; $(\partial/\partial t)$, $(\partial/\partial \phi)$ and 
$(\partial/\partial \psi)$ where $(\partial/\partial \phi)$ and 
$(\partial/\partial \psi)$ are spacelike vectors and $(\partial/\partial t)$ 
is a timelike vector. 
As explained in the last section, 
solutions of vacuum Einstein equations in five-dimensional spacetime with 
three commutable Killing vector fields can be represented by rod structures. 
In this section, we explain how to construct the (diagonal) metric corresponding to 
a given rod structure. 
In a spacetime with a diagonal metric, the directions of the rods are 
 $(\partial/\partial t)$, $(\partial/\partial \phi)$ or
$(\partial/\partial \psi)$. 
Therefore, we suppose that all rods in the seed metric direct to 
$(\partial/\partial t)$, $(\partial/\partial \phi)$ and/or
$(\partial/\partial \psi)$. 
In order to construct asymptotically flat spacetime with black rings (and a black hole),
we prepare the seed metric only with linear mass density $1/2$ and linear mass density $-1/2$. 
Thus, we consider the case where a seed metric has only such a rods. 

\begin{figure}[t]
  \begin{center}
    \includegraphics[keepaspectratio=true,height=30mm]{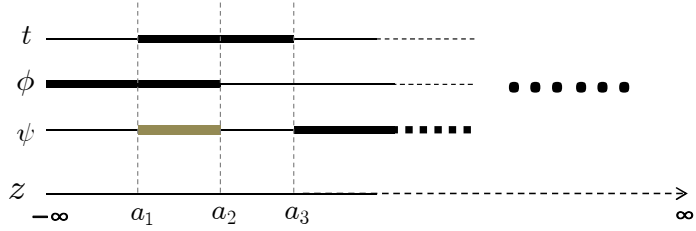}
  \end{center}
  \caption
{Rod structure of example: linear mass density $1/2$ is described as the thick 
black lines. The thick gray line means linear mass density $-1/2$. }
  \label{4-fig2}
\end{figure}  

We explain the method to construct the seed metric from a given rod structure 
with an example described as Fig. \ref{4-fig2}. 
First, we write the minus sign ($-$) in $tt$ component. 
Next, we focus on the left end ($z=-\infty$) of the rod structure. 
We put $\rho^2$ in the numerator of the corresponding component to the existence
of a rod at the left end. 
In the example, $\rho^2$ appears in $\phi\phi$ component. 
Moving to the right (increasing $z$) until the configurations of rods change, 
we construct the soliton corresponding to the changing point. 
If $z=a_1$ there, the constructed soliton is 
\begin{eqnarray}
\mu_1= \sqrt{\rho^2 +(z-a_1)^2}-(z-a_1).
\end{eqnarray}
If a rod with linear mass density $1/2$ ends on some component, we add the soliton 
in the denominator of the corresponding component. 
On the contrary, if a rod with linear mass density $1/2$ starts, the soliton is added 
in the numerator. 
The soliton appears in the numerator of the corresponding component to 
the starting point of a rod with  linear mass density  $-1/2$, while it appears in the 
denominator of the corresponding component to the end point.
Repeating this operation to reach the right end ($z=\infty$), 
the obtained metric has the originally-provided rod structure. 
In the example, the obtained metric is written as 
\begin{eqnarray}
g_{ij}= \mbox{diag}\left(
-\frac{\mu_1\cdots}{\mu_3\cdots},\frac{\rho^2 \cdots}{\mu_2\cdots}, 
\frac{\mu_2\mu_3\cdots}{\mu_1\cdots}
\right).
\end{eqnarray}

It is easy to make sure the obtained metric corresponds to the originally-provided 
rod structure. 
Rods with  linear mass density $1/2$ and rods with  linear mass density $-1/2$ 
mean the corresponding components become $O(\rho^2)$ and $O(\rho^{-2})$ in the 
$\rho\to 0$ limit, respectively. 
In this limit, a soliton behaves $O(\rho^0)$ for $z<a_p$ and $O(\rho^2)$ for $z>a_p$ 
where $a_p$ is the constant parameter of the soliton. 
Adding the soliton $\mu_p$ in numerators is multiplying $O(\rho^2)$ 
only for $z>a_p$, and thus, at the point $z=a_p$ the corresponding 
rod with  linear mass density $1/2$ ends or the corresponding a rod with  linear mass density $-1/2$ starts. 
Putting the soliton $\mu_p$ in denominators bring in the opposite effects.

\section{Solution-generating technique II: B\"{a}cklund transformation}
\subsection{B\"{a}cklund transformation}
In this section we briefly explain the solution-generating technique for the five-dimensional Einstein equations, which is called B\"{a}cklund transformation. This is essentially the technique to generate new solutions of the Ernst equation from a known solution. By using this technique several singly spinning solutions for the five-dimensional spacetime were constructed.\cite{Mishima:2005id,Iguchi:2006tu,Iguchi:2006rd,Iguchi:2007is,Tomizawa:2007mz,Iguchi:2007xs}

We start the analysis from the following form of the metric,
\begin{equation}
ds^2 = e^{-T}\left[
       -e^{S}(dt-\omega d\phi)^2
       +e^{T+2U_1}\rho^2(d\phi)^2 
+e^{2(\gamma+U_1)+T}\left(d\rho^2+dz^2\right) \right]
  +e^{2T}(d\psi)^2.
\label{eq:MBmetric}
\end{equation}
Using this metric form
the Einstein equations are reduced to the following set of equations, 
\begin{eqnarray*}
&&{\bf\rm (i)}\quad
\nabla^2T\, =\, 0,   \\
&&{\bf\rm (ii)}\quad
{
\left\{\begin{array}{ll}
& \partial_{\rho}\gamma_{T}={
  \frac{3}{4}\,\rho\,
  \left[\,(\partial_{\rho}T)^2-(\partial_{z}T)^2\,\right]}\,\ \   \\[3mm]
& \partial_{z}\gamma_T={\displaystyle 
\frac{3}{2}\,\rho\,
  \left[\,\partial_{\rho}T\,\partial_{z}T\,\right],  }
 \end{array}\right. } \\
&&{\bf\rm (iii)}\quad
\nabla^2{\mathcal E}_S=\frac{2}{{\mathcal E}_S+{\bar{\mathcal E}}_S}\,
                    \nabla{\mathcal E}_S\cdot\nabla{\mathcal E}_S , \\  
&&{\bf\rm (iv)}\quad
{
\left\{\begin{array}{ll}
& \partial_{\rho}\gamma_S={\displaystyle
\frac{\rho}{2({\mathcal E}_S+{\bar{\mathcal E}}_S)}\,
  \left(\,\partial_{\rho}{\mathcal E}_S\partial_{\rho}{\bar{\mathcal E}}_S
  -\partial_{z}{\mathcal E}_S\partial_{z}{\bar{\mathcal E}}_S\,
\right)}     \\
& \partial_{z}\gamma_S={\displaystyle
\frac{\rho}{2({\mathcal E}_S+{\bar{\mathcal E}}_S)}\,
  \left(\,\partial_{\rho}{\mathcal E}_S\partial_{z}{\bar{\mathcal E}}_S
  +\partial_{\rho}{\mathcal E}_S\partial_{z}{\bar{\mathcal E}}_S\,
  \right)},
\end{array}\right. } \\
&&{\bf\rm (v)}\quad
\left( \partial_{\rho}\Phi,\,\partial_{z}\Phi \right)
=\rho^{-1}e^{2S}\left( -\partial_{z}\omega,\,\partial_{\rho}\omega \right),  \\
&&{\bf\rm (vi)}\quad 
\gamma=\gamma_S+\gamma_T,   \\
&&{\bf\rm (vii)}\quad 
U_1=-\frac{S+T}{2},
\end{eqnarray*}
where the function $\Phi(\rho,z)$ is defined through the equation (v) and the function 
$\mathcal{E_S}$ is defined by 
$
\,{\mathcal E}_S:=e^{S}+i\,\Phi\,.
$ 
It should be noted that $e^{S}$ and $\Phi$ corresponds to a gravitational potential and a twist potential. The equation (iii) is exactly the same as the Ernst equation in four dimensions \cite{Ernst:1967wx}.
The most nontrivial task to obtain new metrics is to solve 
the equation (iii) because of its nonlinearity. 
We use the method similar to the Neugebauer's 
B\"{a}cklund transformation \cite{Neugebauer:1980} 
or the Hoenselaers-Kinnersley-Xanthopoulos transformation 
\cite{Hoenselaers:1979mk}. 

Following the procedure
given by Castejon-Amenedo and Manko \cite{CastejonAmenedo:1990zz}, for a static seed solution $e^{S^{(0)}}$ a new Ernst potential can be written in the form
\begin{equation}
{\cal E}_S = e^{S^{(0)}}\frac{x(1+ab)+iy(b-a)-(1-ia)(1-ib)}
                         {x(1+ab)+iy(b-a)+(1-ia)(1-ib)}, \nonumber
\end{equation}
where $x$ and $y$ are the prolate-spheroidal coordinates:
$
\,\rho=\sigma\sqrt{x^2-1}\sqrt{1-y^2},\ z=\sigma xy\,
$
with the ranges $1\le x$ and $-1 \le y \le 1$,
and the functions $a$ and $b$ satisfy the following 
simple first-order differential equations 
\begin{eqnarray}
(\ln a)_{,x}&=&\frac{1}{x-y}
[(xy-1)S^{(0)}_{,x}+(1-y^2)S^{(0)}_{,y}], \nonumber \\
(\ln a)_{,y}&=&\frac{1}{x-y}
[-(x^2-1)S^{(0)}_{,x}+(xy-1)S^{(0)}_{,y}], \nonumber\\
(\ln b)_{,x}&=&-\frac{1}{x+y}
\left[(xy+1)S^{(0)}_{,x}+(1-y^2)S^{(0)}_{,y}\right], \nonumber\\
(\ln b)_{,y}&=&-\frac{1}{x+y}
\left[-(x^2-1)S^{(0)}_{,x}+(xy+1)S^{(0)}_{,y}\right]. \nonumber\\
&&    \label{eq:ab}
\end{eqnarray}
The corresponding expressions for the metric functions can be obtained
by using the formulas shown by Ref.~\citen{CastejonAmenedo:1990zz}.  
For the seed, 
\begin{equation}
ds^2 = e^{-T^{(0)}}\left[
       -e^{S^{(0)}}dt^2
       +e^{-S^{(0)}}\rho^2(d\phi)^2 
+e^{2\gamma_{(0)}-S^{(0)}}\left(d\rho^2+dz^2\right) \right]
  +e^{2T^{(0)}}(d\psi)^2,
\label{eq:backseed}
\end{equation}
 a new solution is given by
\begin{eqnarray}
ds^2&=&-e^{S^{(0)}-T^{(0)}}\frac{A}{B}\biggl[dt-\biggl(2\sigma e^{-S^{(0)}}\frac{C}{A}+C_1\biggr)d\phi\biggr]^2
\nonumber\\    & &
+\frac{B}{A}e^{-S^{(0)}-T^{(0)}}\sigma^2(x^2-1)(1-y^2)d\phi^2
\nonumber\\    & &
+e^{2T^{(0)}}d\psi^2
+C_2 e^{2\gamma'-S^{(0)}-T^{(0)}} {B} \sigma^2 \frac{x^2-y^2}{x^2-1}\biggl(\frac{dx^2}{x^2-1}+\frac{dy^2}{1-y^2}\biggr), \label{eq:sol}
\end{eqnarray} 
where $C_1$ and $C_2$ are constants and $A$, $B$ and $C$ are defined by
\begin{eqnarray}
A &:=& (x^2-1)(1+ab)^2-(1-y^2)(b-a)^2,\nonumber\\
B &:=& [(x+1)+(x-1)ab]^2+[(1+y)a+(1-y)b]^2,\nonumber\\
C &:=& (x^2-1)(1+ab)[(1-y)b-(1+y)a]
  \nonumber\\&&+(1-y^2)(b-a)[(x+1)-(x-1)ab].\label{eq:A-C}
\end{eqnarray}  
The function $\gamma'$ in Eq. (\ref{eq:sol}) is a $\gamma$ function corresponding to the static metric,
\begin{eqnarray}
ds^2 &=& e^{-T^{(0)}}\left[
       -e^{2U^{\mbox{\tiny (BH)}}_0+S^{(0)}}(dx^0)^2
       +e^{-2U^{\mbox{\tiny (BH)}}_0-S^{(0)}}\rho^2(d\phi)^2  \right. \nonumber \\ && \left.
   +e^{2(\gamma'-U^{\mbox{\tiny (BH)}}_0)-S^{(0)}}\left(d\rho^2+dz^2\right) \right]
  +e^{2T^{(0)}}(d\psi)^2, \nonumber \\ \label{static_5}
\end{eqnarray}
where ${\displaystyle U_{0}^{\mbox{\tiny (BH)}}=\frac{1}{2}\ln\left( \frac{x-1}{x+1} \right)}$. 

Next we consider the solutions 
of the differential equations (\ref{eq:ab}). 
At first, we examine the case of a typical seed function 
\begin{equation}
S^{(0)}=\frac{1}{2}\ln\left[\,R_{d}+(z-d)\,\right], 
\label{seed_g}
\end{equation}
where  $R_{d}=\sqrt{\rho^2+(z-d)^2}$. 
The general seed function is composed of seed functions of this form.
Note that the above function $S^{(0)}$ is a Newtonian potential 
whose source is 
a semi-infinite thin rod \cite{Emparan:2001wk}.

In this case
we can confirm that the following $a$ and $b$ satisfy the 
differential equations (\ref{eq:ab}),
\begin{equation}
a=l_{\sigma}^{-1}e^{2\phi_{d,\sigma}}\ ,\ \ \ 
b=-l_{-\sigma}e^{-2\phi_{d,-\sigma}}\ ,\label{ab_phi}
\end{equation}
where
\begin{equation}
\phi_{d,c}=\frac{1}{2}
 \ln\left[\,e^{-\tilde{U}_{d}}\left(e^{2U_c}+e^{2\tilde{U}_{d}}\right)\,\right].
 \label{phi_i}
\end{equation}
Here the functions $\tilde{U}_{d}$ and $U_{c}$ are defined as $\tilde{U}_{d}:=\frac{1}{2}\ln\left[\,R_{d}+(z-d)\,\right]$ and 
$U_{c}:=\frac{1}{2}\ln\left[\,R_{c}-(z-c)\,\right]$.
Because of the linearity of the differential equations (\ref{eq:ab})
for $S^{(0)}$,
we can easily obtain $a$ and $b$ which correspond to a general seed function 
if it is a linear combination of (\ref{seed_g}).

The function $\gamma'$ is defined from the static metric (\ref{static_5}), 
so that $\gamma'$ obeys the following equations,
\begin{equation}
\partial_{\rho}\gamma'=
\frac{1}{4}\rho\left[(\partial_{\rho}S')^2-(\partial_{z}S')^2\right]
 +\frac{3}{4}\rho\left[(\partial_{\rho}T')^2-(\partial_{z}T')^2\right],
 \label{eq:drho_gamma'}
\end{equation}
\begin{equation}
\partial_{z}\gamma'=
\frac{1}{2}\rho\left[\partial_{\rho}S'\partial_{z}S'\right]
 +\frac{3}{2}\rho\left[\partial_{\rho}T'\partial_{z}T'\right],
 \label{eq:dz_gamma'}
\end{equation}
where the first terms are contributions from Eq. (iv) and the 
second terms come from Eq. (ii).
Here the functions $S'$ and $T'$ can be read out from Eq. (\ref{static_5}) as
\begin{eqnarray}
S'&=&2\,U^{(BH)}_0+S^{(0)},  \label{eq:S'} \\
T'&=&T^{(0)}. \label{eq:T'}
\end{eqnarray}
To integrate these equations we can use the following fact 
that the partial differential equations
\begin{eqnarray}
\partial_{\rho}\gamma'_{cd}
    &=&\rho\left[\partial_{\rho}\tilde{U}_{c}\partial_{\rho}\tilde{U}_{d}
            -\partial_{z}\tilde{U}_{c}\partial_{z}\tilde{U}_{d}\right],  \label{drho_gm}\\
\partial_{z}\gamma'_{cd}
    &=&\rho\left[\partial_{\rho}\tilde{U}_{c}\partial_{z}\tilde{U}_{d}
           +\partial_{\rho}\tilde{U}_{d}\partial_{z}\tilde{U}_{c}\right], \label{dz_gm}
\end{eqnarray}
have the following solution, 
\begin{equation}
\gamma'_{cd}=\frac{1}{2}\tilde{U}_{c}+\frac{1}{2}\tilde{U}_{d}-\frac{1}{4}\ln Y_{cd}, \label{gam'}
\end{equation}
where $Y_{cd}:=R_cR_d+(z-c)(z-d)+\rho^2$. The general solution of $\gamma'$
is given by the linear combination of the functions $\gamma'_{cd}$.

\subsection{Relation between two solution-generating methods}
To construct five-dimensional solutions, we have been mainly using two different solitonic solution-generating techniques, the inverse scattering method and the B\"{a}cklund transformation.
Both methods successfully constructed several important solutions. 
Also, these two methods will be used to find new higher-dimensional black hole solutions.
Note that while the solutions with single angular momentum component can be constructed by the both methods, the solutions with double angular momentum components can be constructed only by the inverse scattering method.

It is important to consider the relation between solutions generated by these two methods.
The relation between the four-dimensional solutions generated by these two methods was investigated in Refs.~\citen{Gurses:1985hf,Gurses:1983up,Cosgrove:1980fx,Stephani:2003tm}. 
Recently the relation between the five-dimensional solutions was also investigated by Tomizawa et al.\cite{Tomizawa:2006jz}
The singly spinning five-dimensional solutions found or reconstructed by the B\"{a}cklund transformation can also be constructed by the inverse scattering methods.
The solutions obtained by B\"{a}cklund transformation are 2-soliton solutions whose seed metrics are diagonal.
It was shown that the 2-soliton solutions generated by the inverse scattering method coincide with ones generated 
by the B\"{a}cklund transformation for the general diagonal seed.
In fact, the five-dimensional black ring with $S^2$ rotation was constructed from the Minkowski spacetime by using both methods.\cite{Mishima:2005id,Tomizawa:2005wv}
Also it was shown that the black ring with $S^1$ rotation can be constructed from the Euclidean $C$-metric by using both methods.\cite{Iguchi:2006rd,Tomizawa:2006vp}

It has been often used the inverse scattering method that was modified by Pomeransky to construct the five-dimensional solutions.\cite{Pomeransky:2005sj}
In single-rotational case we can easily transform the procedure of the original inverse scattering method into the modified method. This means that we can investigate the relation between solutions obtained by the B\"{a}cklund transformation and the modified inverse scattering method through the mediation of the original inverse scattering method.

\section{Single horizon solution}
\label{4-Sec:single}
Previously five-dimensional black hole solutions which have only one event horizon were found without the use of solitonic solution-generating techniques. 
These solutions were reconstructed and generalized by using the solitonic techniques.
In this section we briefly explain the ways to construct the solutions of five-dimensional Myers-Perry black hole and black rings.

\subsection{Myers-Perry black hole I}
The Myers-Perry black hole solution with single angular momentum component was regenerated from the Minkowski seed by using the solitonic techniques. 
This solution is obtained as a limit solution of $S^2$ rotating black ring without the inside space of the ring.\cite{Mishima:2005id,Tomizawa:2005wv}
Also it can be directly derived from the same seed of $S^2$ rotating black ring by adding solitons at appropriate positions.

\begin{figure}[t]
  \begin{center}
    \includegraphics[keepaspectratio=true,height=30mm]{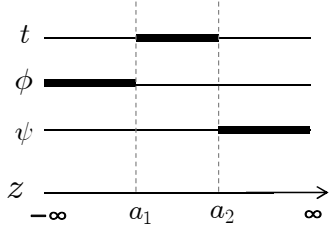}
  \end{center}
  \caption{Rod structure for the seed metric $g^{(0)}$ of Myers-Perry black hole I. Solid lines corresponding to sources of  linear mass density $1/2$.
  }
  \label{fig:rod_seed_MP1}
\end{figure}

The doubly spinning solution of Myers-Perry black hole was regenerated by using the inverse scattering method.\cite{Pomeransky:2005sj}
To obtain the seed metric of the solution we start from the Schwarzschild-Tangherlini black hole
\begin{equation}
 g_{ij}^{(0)} = {\mbox{diag}} \left( - \frac{\mu_1}{\mu_2}, \frac{\rho^2}{\mu_1}, \mu_2 \right) 
\end{equation}
and
\begin{equation}
 f^{(0)} = \mu_2 \frac{\rho^2 + \mu_1 \mu_2}{(\rho^2 + \mu_1^2)(\rho^2 + \mu_2^2)}.
\end{equation}
The rod structure of this metric is shown in Fig. \ref{fig:rod_seed_MP1}.
We divide the $tt$ component of $g^{(0)}$ by $-\frac{\mu_2^2}{\rho^2}$ and $-\frac{\rho^2}{\mu_1^2}$ by which we effectively remove an anti-soliton at $z=a_1$ and a soliton at $z=a_2$ from $g^{(0)}$.
We denote the obtained seed metric as $g'$, 
\begin{equation}
 {g'}_{ij}^{(0)} = {\mbox{diag}} \left( - \frac{\mu_2}{\mu_1}, \frac{\rho^2}{\mu_1}, \mu_2 \right). 
\end{equation}
For the convenience, we use the seed metric which is rescaled by a factor of $-\frac{\mu_1}{\mu_2}$,
\begin{equation}
 \tilde{g}_{ij}^{(0)} = {\mbox{diag}} \left( 1 , \bar{\mu}_2 , -\mu_1 \right).
\end{equation}
The corresponding generating matrix $\tilde{\Psi}^{(0)}$ which satisfies Eqs. (\ref{4-Psi1}) and (\ref{4-Psi}) is given by
\begin{equation}
 \tilde{\Psi}^{(0)} = \mbox{diag} \left( 1, (\bar{\mu_2} - \lambda), -(\mu_1 - \lambda) \right).
\end{equation}
We perform a 2-soliton transformation with $\tilde{g}^{(0)}$ where an anti-soliton is placed at $z=a_1$ with BZ vector $m_0^{(1)} = (1, b, 0)$ and a soliton at $z=a_2$ with $m_0^{(2)} = (1,0,c)$. 
We denote the resulting metric as $\tilde{g}$ and rescale it to find the metric $g = -\frac{\mu_2}{\mu_1} \tilde{g}$.
There is a freedom to make linear transformation in the space of coordinates $t$, $\phi$ and $\psi$. 
For example, we perform a following linear coordinate transformation
\begin{eqnarray}
 t & = & t^{\mbox{\scriptsize new}} + 2 b (a_2-a_1) \phi^{\mbox{\scriptsize new}} + 2 c (a_2 -a_1) \psi^{\mbox{\scriptsize new}} \\
 \phi & = & 2 (a_1 - a_2) \phi^{\mbox{\scriptsize new}} + bc \psi^{\mbox{\scriptsize new}} \\
 \psi & = & 2 (a_1 - a_2) \psi^{\mbox{\scriptsize new}} + bc \phi^{\mbox{\scriptsize new}} 
\end{eqnarray}
to remove global rotations of the solution.
The conformal factor of two-dimensional part $f$ is given by
\begin{equation}
 f = 4(a_1-a_2)^2 f^{(0)} \frac{\det \Gamma}{\det \Gamma^{(0)}}
\end{equation}
where $\Gamma$ is found in the process of constructing $g$ and $\Gamma^{(0)} = \Gamma|_{b=c=0} $.

\subsection{Myers-Perry black hole II}
\begin{figure}[t]
  \begin{center}
    \includegraphics[keepaspectratio=true,height=30mm]{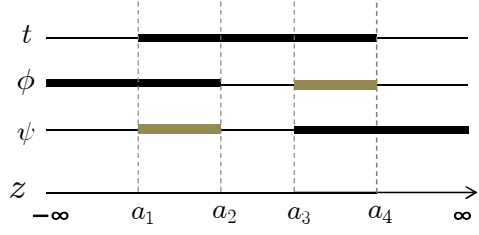}
  \end{center}
  \caption{Rod structure for the seed metric $g^{(0)}$ of Myers-Perry black hole II. 
  Black lines corresponding to sources of  linear mass density  $1/2$ and gray lines to 
   linear mass density  $-1/2$.}
    \label{fig:rod_seed_MP2}
\end{figure}

There is another way to construct the doubly spinning solution of Myers-Perry black hole by using the inverse scattering method. 
We start from the following seed metric
\begin{equation}
 g_{ij}^{(0)} = {\mbox{diag}} \left( - \frac{\mu_1}{\mu_4},\frac{\rho^2 \mu_4}{\mu_2 \mu_3}, \frac{\mu_2 \mu_3}{\mu_1} \right) 
\end{equation}
and
\begin{equation}
 f^{(0)} = \frac{\mu_2 \mu_3}{\mu_1} \frac{W_{12} W_{13} W_{14} W_{24} W_{34}}{W_{23}^2 W_{11} W_{22} W_{33} W_{44}}
\end{equation}
where $W_{pq} =  \rho^2 + \mu_p \mu_q$.
The rod structure of this metric is shown in Fig. \ref{fig:rod_seed_MP2}.
At first we remove a soliton at $z=a_1$ and an anti-soliton at $z=a_2$ with trivial BZ vectors $(1,0,0)$ as
\begin{equation}
 {g'}_{ij}^{(0)} = {\mbox{diag}} \left( - \frac{\mu_4}{\mu_1},\frac{\rho^2 \mu_4}{\mu_2 \mu_3}, \frac{\mu_2 \mu_3}{\mu_1} \right). 
\end{equation}
For the convenience, the seed metric is rescaled by a factor of $-\frac{\mu_1}{\mu_4}$,
\begin{equation}
 \tilde{g}_{ij}^{(0)} = {\mbox{diag}} \left( 1 , \frac{\bar{\mu}_2 \mu_1}{\mu_3} , - \frac{\mu_2 \mu_3}{\mu_4} \right).
\end{equation}
After construction of the generating matrix as
\begin{equation}
 \tilde{\Psi}^{(0)} = {\mbox{diag}} \left( 1 , \frac{(\bar{\mu}_2 - \lambda)(\mu_1 - \lambda)}{\mu_3 - \lambda} , - \frac{(\mu_2 - \lambda)(\mu_3 - \lambda)}{\mu_4 - \lambda} \right),
\end{equation}
we perform a 2-soliton transformation with $\tilde{g}^{(0)}$ where an  anti-soliton is placed at $z=a_1$ with BZ vector $m_0^{(1)} = (1, 0, c)$ and a soliton at $z=a_4$ with $m_0^{(2)} = (1,b,0)$. Finally we rescale the metric $\tilde{g}$ to find the metric $g = -\frac{\mu_4}{\mu_1} \tilde{g}$. 
The conformal factor $f$ is given by
\begin{equation}
 f = f^{(0)} \frac{\det \Gamma}{\det \Gamma^{(0)}},
\end{equation}
where $\Gamma^{(0)} = \Gamma|_{b=c=0}$.
The BZ parameters should be determined to remove the singularities on the axis as
\begin{equation}
 b = \pm \sqrt{\frac{2 (a_4 - a_1)(a_4 - a_2)}{(a_4 - a_3)}}, ~~~ c = \pm \sqrt{\frac{2 (a_2 - a_1)(a_3 - a_1)}{(a_4 - a_1)}}. 
\end{equation}
The obtained metric $(g,f)$ already has the same asymptotic form as the Minkowski spacetime. Therefore we do not
need to perform a linear transformation.

\subsection{Black ring}
The five-dimensional black ring has an event horizon whose topology is $S^1 \times S^2$. The black ring has to rotate along the $S^1$ direction to avoid a conical singularity in the inside space of the ring. The original solution of black ring which has only one angular momentum component along the $S^1$ direction was found without use of solitonic techniques.\cite{Emparan:2001wn}
 The black ring solutions which have angular momentum component along the $S^2$ direction were constructed by using the solitonic solution-generating techniques.

\subsubsection{Black ring with $S^2$ rotation} 
\begin{figure}[t]
  \begin{center}
    \includegraphics[keepaspectratio=true,height=30mm]{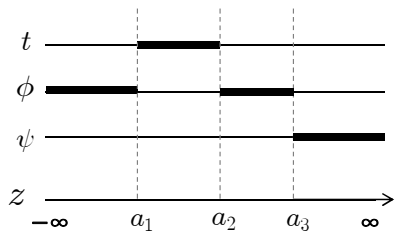}
  \end{center}
  \caption{Rod structure for the seed metric $g^{(0)}$ of $S^2$ rotating 
  black ring. Solid lines corresponding to sources of  linear mass density $1/2$.}
    \label{fig:rod_seed_S2BR}
\end{figure}

The solution of black ring with $S^2$ rotation was constructed by using the B\"{a}cklund transformation\cite{Mishima:2005id,Iguchi:2006tu} and reconstructed by using the inverse scattering method.\cite{Tomizawa:2005wv} Also the same solution was found by the educational guess work.\cite{Figueras:2005zp}
The seed metric of $S^2$ rotating black ring is the five-dimensional Minkowski spacetime when we use the B\"{a}cklund transformation.
While we can construct the same solution from the same Minkowski seed by using the inverse scattering method,
it is convenient to start from the metric of static black ring,
\begin{equation}
  g_{ij}^{(0)} = {\mbox{diag}} \left( - \frac{\mu_1}{\mu_2}, \frac{\rho^2 \mu_2}{\mu_1 \mu_3}, {\mu_3} \right)
\end{equation}
and
\begin{equation}
 f^{(0)} = \mu_3 \frac{W_{12}^2 W_{23}}{W_{13}\prod _{i=1}^3 W_{ii} } .
\end{equation}
The rod structure of this seed metric is shown in Fig. \ref{fig:rod_seed_S2BR}.
At first we remove an anti-soliton at $z=a_1$ and a soliton at $z=a_2$ with trivial BZ vectors and next rescale it by a factor $-\frac{\mu_2}{\mu_1}$. The resulting metric is
\begin{equation}
 \tilde{g}_{ij}^{(0)} = {\mbox{diag}} \left( 1 , \bar{\mu}_3 , -\frac{\mu_1 \mu_3}{\mu_2} \right).
\end{equation}
After constructing a generating matrix from this metric, 
\begin{equation}
 \tilde{\Psi}^{(0)} = {\mbox{diag}} \left( 1 , \bar{\mu}_3 - \lambda , - \frac{(\mu_1 - \lambda)(\mu_3 - \lambda)}{\mu_2 - \lambda} \right),
\end{equation}
we perform a 2-soliton transformation to obtain the metric $\tilde{g}$ by adding an anti-soliton at $z=a_1$ with a BZ vector $(1,b_1,0)$ and a soliton at $z=a_2$ with a BZ vector $(1,b_2,0)$ to the seed $\tilde{g}_{ij}^{(0)}$. 
Finally we rescale the metric $\tilde{g}$ to find the metric $g=-\frac{\mu_2}{\mu_1} \tilde{g}$.
The conformal factor $f$ is given by
\begin{equation}
 f = \frac{4(a_1-a_3)^2}{(2(a_1-a_3)-b_1 b_2)^2} f^{(0)} \frac{\det \Gamma}{\det \Gamma^{(0)}},
\end{equation}
where $\Gamma^{(0)} = \Gamma|_{b_1=b_2=0}$.
In order for the metric to be asymptotically Minkowski, we need perform an appropriate linear transformation of the coordinates $t$ and $\phi$,
\begin{equation}
 t = t^{\mbox{\scriptsize new}} - \frac{2(a_1-a_2) b_1}{2(a_1 - a_3) - b_1 b_2} \phi^{\mbox{\scriptsize new}}, ~~~ \phi = \phi^{\mbox{\scriptsize new}}.
\end{equation}
In general the solutions has some undesiered features.
When the parameters satisfy the following equation
\begin{equation}
 2(a_2 -a_3) b_1 + 2(a_1 - a_3) b_2 - b_1 b_2 (b_1 + b_2)=0,
\end{equation}
the finite spacelike rod between $z=a_2$ and $z=a_3$ corresponds with the $\phi$ axis where $g_{\phi\phi}=0$.
Even in this case, we can not remove conical singularities in the inside space of the ring.

\subsubsection{Black ring with $S^1$ rotation}
\label{sec:S1_black_ring}
\begin{figure}[t]
  \begin{center}
    \includegraphics[keepaspectratio=true,height=30mm]{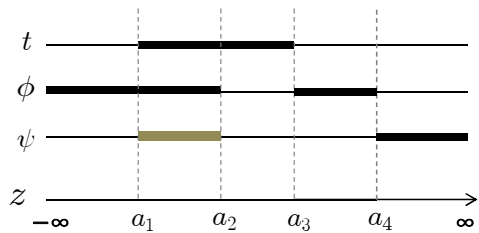}
  \end{center}
  \caption{Rod structure for the seed metric $g^{(0)}$ of $S^1$ rotating black ring.
   Black lines corresponding to sources of  linear mass density $1/2$ and 
   gray lines to  linear mass density  $-1/2$.}
     \label{fig:rod_seed_S1BR}
\end{figure}

The black ring with an angular momentum component along $S^1$ direction was reconstructed by the B\"{a}cklund transformation\cite{Iguchi:2006rd}
 and the inverse scattering method\cite{Tomizawa:2006vp}
 starting from a Euclidean C-metric as a seed. In these analyses, a 2-soliton transformation was used to derive the solution.
We can also construct the $S^1$ rotating black ring by a 1-soliton transformation from another seed.\cite{Pomeransky:2006bd,Emparan:2008eg}
To do so, we start from the following diagonal metric,
\begin{equation}
  g_{ij}^{(0)} = {\mbox{diag}} \left( - \frac{\mu_1}{\mu_3}, \frac{\rho^2 \mu_3}{\mu_2 \mu_4}, \frac{\mu_2 \mu_4}{\mu_1} \right). 
\end{equation}
The conformal factor of the seed is given by
\begin{equation}
 f^{(0)} = \frac{\mu_2 \mu_4}{\mu_1} \frac{W_{12}^2 W_{13} W_{15} W_{16} W_{23} W_{25} W_{34}^2 W_{36} W_{45}^2 W_{56}}{W_{14} W_{24} W_{26}^2 W_{35}^2 W_{46} \prod _{i=1}^6 W_{ii} }.
\end{equation}
The rod structure of this metric is shown in Fig. \ref{fig:rod_seed_S1BR}. 
We need a negative density rod between $z=a_1$ and $a_2$ to obtain the $S^1$ rotating black ring.
We remove an anti-soliton at $z=a_1$ with trivial BZ vector and multiply the factor $\frac{\mu_1}{\rho^2}$ to the metric.
The resulting metric is denoted as $\tilde{g}^{(0)}$,
\begin{equation}
  \tilde{g}_{ij}^{(0)} = {\mbox{diag}} \left( \frac{1}{\mu_3} , \frac{\bar{\mu}_2 \mu_3}{\bar{\mu}_1 \mu_4} , -\frac{\mu_2}{\bar{\mu}_4} \right).
\end{equation}
Then we construct a generating matrix from $\tilde{g}^{(0)}$ as
\begin{equation}
 \tilde{\Psi}^{(0)} = {\mbox{diag}} \left( \frac{1}{\mu_3 - \lambda} , \frac{(\bar{\mu}_2 - \lambda)(\mu_3 - \lambda)}{(\bar{\mu}_1 - \lambda)(\mu_4 - \lambda)} , - \frac{(\mu_2 - \lambda)}{\bar{\mu}_4 - \lambda} \right),
\end{equation}
Next we readd an anti-soliton with a nontrivial BZ vector $(1,0,c)$ and obtain a metric $\tilde{g}$. Finally we rescale the metric $\tilde{g}$ to obtain the metric $g = \frac{\rho^2}{\mu_1} \tilde{g}$.
The conformal factor $f$ is given by
\begin{equation}
 f = f^{(0)} \frac{\det \Gamma}{\det \Gamma^{(0)}}
\end{equation}
where $\Gamma^{(0)} = \Gamma|_{c=0}$.
To remove the singular behavior on the axis we set the BZ parameter as
\begin{equation}
 c = \sqrt{\frac{2(a_2 -a_1)(a_3- a_1)}{a_4 -a_1}}.
\end{equation}
In this case we do not need to perform a linear transformation of the coordinates.
In general the solution has a conical singularity on the plane of the ring.
The balance condition to avoid a conical singularity is given by, for example,
\begin{equation}
 a_1 = \frac{a_2^2 - 2 a_2 a_4 + a_3 a_4}{a_3 - a_4}.
\end{equation}

\subsubsection{Doubly spinning black ring}
The doubly spinning black ring solution was found by using the inverse scattering method.\cite{Pomeransky:2006bd}
The seed metric of this solution was obtained by removing a pair of solitons with trivial BZ vectors from the Emparan-Reall black ring which is rotating along $S^1$ direction. Therefore the seed metric is not diagonal. 
To rotate the Emparan-Real black ring along  the $S^2$ direction we readd the same pair of solitons with nontrivial BZ vectors to the nondiagonal seed.
The most disturbing difficulty is to find a generating matrix field $\Psi^{(0)}$ which corresponds to the nondiagonal metric of Emparan-Reall black ring.
Fortunately, it is known that if we construct the $S^1$ rotating black ring via a one-soliton transformation as in \ref{sec:S1_black_ring}, this matrix field automatically 
is given.
It is worth to note that the regular black ring with two angular momenta comes from the seed solution, which is obtained from the regular $S^1$ rotating black ring without conical singularity.

\section{Solution of disconnected horizons}
\label{4-Sec:multi}
In five dimensions, in addition to the solutions which have only one event horizon component, there exist solutions which have more than one disconnected event horizons.
These solutions were obtained by using solitonic techniques, the B\"{a}cklund transformation and the inverse scattering method.
In the following subsections we will briefly explain the ways to construct these solutions and the physical features of them.

\subsection{Black Saturn}
\begin{figure}[t]
  \begin{center}
    \includegraphics[keepaspectratio=true,height=30mm]{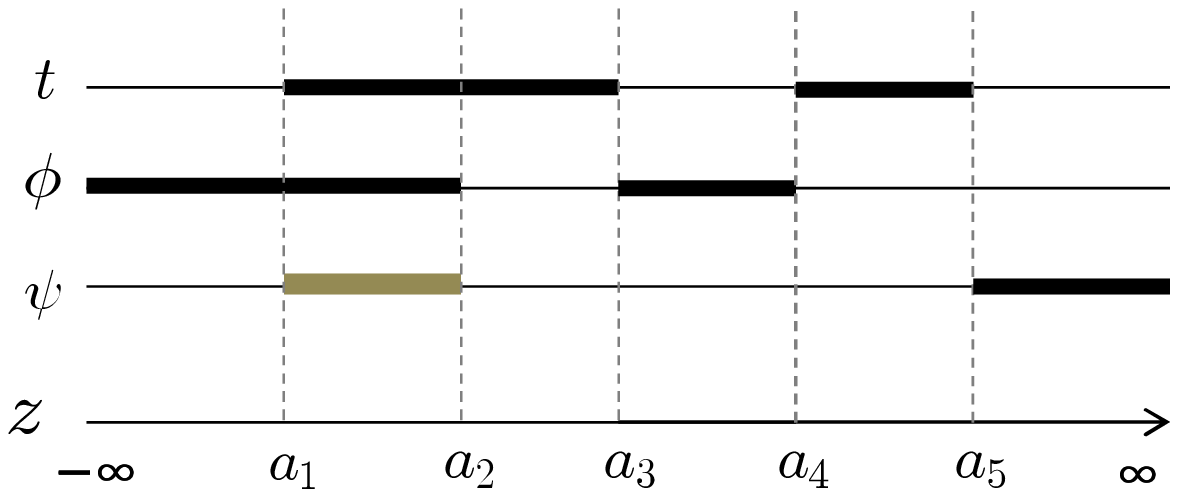}
  \end{center}
  \caption{Rod structure for the seed metric $g^{(0)}$ of black Saturn. 
  Black lines corresponding to sources of  linear mass density  $1/2$ and 
  gray lines to  linear mass density  $-1/2$.}
    \label{fig:rod_seed_BS}
\end{figure}

One of the example of black hole solutions with disconnected horizons is the black Saturn solution,\cite{Elvang:2007rd}
 in which a spherical black hole is surrounded by a black ring.
This solution was constructed by using the inverse scattering method starting from the following seed metric
\begin{equation}
 g_{ij}^{(0)} = {\mbox{diag}} \left( - \frac{\mu_1 \mu_4}{\mu_3 \mu_5}, \frac{\rho^2 \mu_3}{\mu_2 \mu_4}, \frac{\mu_1 \mu_5}{\mu_2} \right)
\end{equation}
and
\begin{equation}
 f^{(0)} = \frac{k^2 \mu _2 \mu _5}{\mu _1}\text{  }\frac{W_{12}W_{13} W_{15}{}^2 W_{23} W_{34}{}^2 W_{45}}{W_{14} W_{24}W_{25} W_{35} \prod _{i=1}^5 W_{ii}}.
\end{equation}
The rod structure of this metric is shown in Fig. \ref{fig:rod_seed_BS}. 
At first we remove anti-solitons at $z=a_1$ and $a_4$ and a soliton at $z=a_5$ with trivial BZ vectors and rescale it by a factor of $\frac{\mu_1 \mu_4}{\rho^2 \mu_2}$ to find
\begin{equation}
   \tilde{g}_{ij}^{(0)} = {\mbox{diag}} \left( \frac{1}{\mu_3} , \frac{\mu_1 \mu_3}{\mu_2 \mu_5} , -\frac{\mu_4}{\bar{\mu}_2} \right).
\end{equation}
The generating matrix is given by
\begin{equation}
 \tilde{\Psi}^{(0)} = {\mbox{diag}} \left( \frac{1}{\mu_3 - \lambda} , \frac{(\mu_1 - \lambda)(\mu_3 - \lambda)}{(\mu_2 - \lambda)(\mu_5 - \lambda)} , - \frac{(\mu_4 - \lambda)}{\bar{\mu}_2 - \lambda} \right).
\end{equation}
Next we perform a 3-soliton transformation with $\tilde{g}^{(0)}$ as seed: Add an anti-soliton at $z=a_1$ with BZ vector $(1,0,c_1)$ and an anti-soliton at $z=a_4$ with $(1,b,0)$ and a soliton at $z=a_5$ with $(1,0,c_2)$. We denote the resulting metric $\tilde{g}$.
Finally we rescale $\tilde{g}$ to find a metric $g=\frac{\rho^2 \mu_5}{\mu_1 \mu_4} \tilde{g}$.
The metric factor $f$ is obtained as
\begin{equation}
 f = f^{(0)} \frac{\det \Gamma}{\det \Gamma^{(0)}}
\end{equation}
where $\Gamma^{(0)} = \Gamma|_{b=c_1=c_2=0} $.

The physical feature of black Saturn was investigated only for the case of single spin solution in which $b=0$.
One of the most important properties of multi-horizon solutions is continuous non-uniqueness of the solution.
To show this property, the phase diagram of the black Saturn was investigated in Refs.~\citen{Elvang:2007rd,Elvang:2007hg}. 
The plot of random sets of points in the phase diagram showed that the black Saturn covers the wide region of the phase diagram. 
This results can be understood by the analysis based on the thin and long ring approximation in which the black Saturn can
be modeled as a simple superposition of an Myers-Perry black hole and a very thin black ring. \cite{Elvang:2007hg}
It was argued that the configurations that approach maximal entropy for fixed mass and angular momentum are black Saturns with a nearly static black hole and a very thin black ring.

Another important property of multi-horizon solutions is the existence of thermodynamical equilibrium configuration in which the horizons have the same temperature and angular velocity. \cite{Elvang:2007hg}
Requiring equal temperatures and angular velocities for the central black hole and the black ring imposes two conditions
on the parameters. This removes entirely the continuous non-uniqueness, leaving at most
discrete degeneracies. Black Saturns in thermodynamical equilibrium thus form a curve in
the phase diagram.
The black Saturn in thermodynamical equilibrium is not thermodynamically stable because the entropy of it is not the maximum.
It has been shown that metastablity occurs when the dimensionless
total angular momentum lies in a narrow window $0.85483 < j^2 < 0.85494$ of the thin ring
branch.\cite{Evslin:2008py}

%

\subsection{Black di-ring}
\subsubsection{Construction of black di-ring}

\begin{figure}[t]
  \begin{center}
    \includegraphics[keepaspectratio=true,height=30mm]{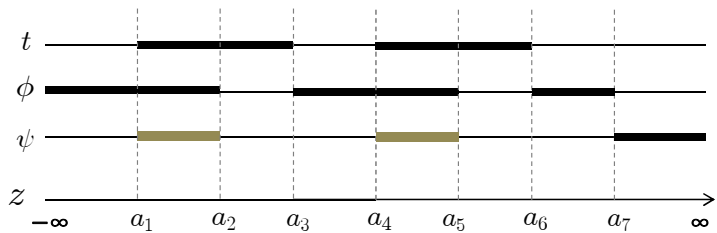}
  \end{center}
  \caption{Rod structure for the seed metric black di-ring which is used in the 
  analysis by Evslin and Krishnan. Black lines corresponding to sources of  linear mass density  $1/2$ and gray lines to  linear mass density  $-1/2$.}
    \label{fig:rod_seed_DR2}
\end{figure}

We can also construct a multi-ring configuration by using solitonic solution-generating techniques.
The first example of multi-ring solution is a black di-ring which has two concentric black rings rotating on the same plane.
The black di-ring solutions were first constructed by using the B\"{a}cklund transformation\cite{Iguchi:2007is} and reconstructed by the inverse scattering method.\cite{Evslin:2007fv}
The representations of these two solutions are very different. One of the reason for this difference is that these two representations are obtained by solitonic techniques starting from different seed metrics. 

In this subsection we derive a representation of black di-ring by using the inverse scattering method starting from a seed metric which is different from the one used in the analysis of Evslin and Krishnan.
This representation of black di-ring directly corresponds to the one obtained by B\"{a}cklund transformation because the seed metrics are essentially same.
The representation obtained by Evslin and Krishnan is presented in Chap. 2, which is derived from the seed metric whose rod structure is given in Fig. \ref{fig:rod_seed_DR2}. 
Considering the relations between these solutions, it was shown that the different solution sets of di-rings which are obtained by the B\"{a}cklund transformation and the inverse scattering method are completely equivalent.\cite{Iguchi:2010pe} 

\begin{figure}[t]
  \begin{center}
    \includegraphics[keepaspectratio=true,height=30mm]{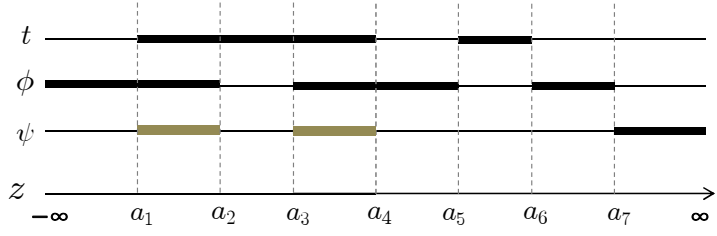}
  \end{center}
  \caption{Rod structure for the seed metric $g^{(0)}$ of black di-ring. 
  Black lines corresponding to sources of  linear mass density  $1/2$ 
  and gray lines to  linear mass density  $-1/2$.}
    \label{fig:rod_seed_DR}
\end{figure}

The seed solution is represented by the rod structure given in Fig. \ref{fig:rod_seed_DR}.
The corresponding seed metric is given by 
\begin{equation}
 g^{(0)}_{ij} = \mbox{diag} \left( -\frac{\mu_1 \mu_5}{\mu_4 \mu_6} , \frac{\rho^2 \mu_3 \mu_6}{\mu_2 \mu_5 \mu_7} , \frac{\mu_2 \mu_4 \mu_7}{\mu_1 \mu_3} \right)
\end{equation}
and
\begin{equation}
f^{(0)} = \frac{k^2 \mu _2\mu _4\mu _7}{\mu _1\mu _3}\text{  }\frac{W_{12}W_{14}{}^2 W_{16}W_{17}W_{23}{}^2 W_{26}W_{34}W_{35}W_{37}{}^2 W_{45}W_{56}{}^2 W_{67}}{W_{13}W_{15}W_{24}W_{25}W_{27}{}^2 W_{36}W_{46}W_{47}W_{57}\prod _{i=1}^7 W_{ii}}.
\end{equation}

\begin{figure}[t]
  \begin{center}
    \includegraphics[keepaspectratio=true,height=30mm]{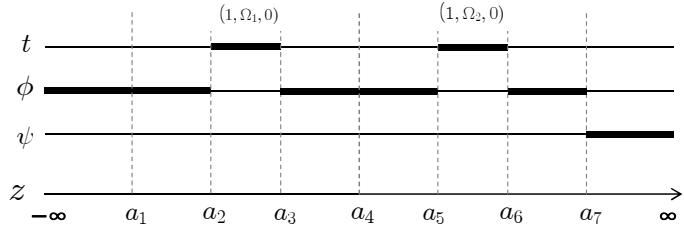}
  \end{center}
  \caption{Rod structure of the regular black di-ring. Direction vectors are presented over each rod. Singularities at $z=a_1$ and $z=a_4$ are removed by setting the BZ parameters $c_1$ and $c_2$ correctly. }
    \label{fig:rod_diring}
\end{figure}

The solution of black di-ring is constructed as follows:
\begin{enumerate}
\item
Perform two 1-soliton transformations on the seed solution: 
Remove an anti-soliton at $z=a_1$ with trivial BZ vector (1,0,0) and a soliton at $z=a_4$ with trivial BZ vector (1,0,0).
The result is the metric matrix
\begin{equation}
 {g'}^{(0)} = \mbox{diag} \left( -\frac{\mu_4 \mu_5}{\mu_1 \mu_6} , \frac{\rho^2 \mu_3 \mu_6}{\mu_2 \mu_5 \mu_7}  , \frac{\mu_2 \mu_4 \mu_7}{\mu_1 \mu_3} \right).
\end{equation}
\item
Rescale ${g'}^{(0)}$ by a factor of $-\displaystyle\frac{\mu_1}{\mu_4}$ to find
\begin{equation}
 \tilde{g}^{(0)} = \mbox{diag} \left( \frac{\mu_5}{\mu_6}  , \frac{\mu_1 \bar{\mu}_2 \mu_3 \mu_6}{\mu_4 \mu_5 \mu_7} ,- \frac{\mu_2 \mu_7}{\mu_3} \right),
\end{equation}
where $\bar{\mu}_2 = -\frac{\rho^2}{\mu_2}$.
This will be the seed for the next solution transformation.
\item
The generating matrix can be found from $\tilde{g}^{(0)}$. It is
\begin{eqnarray}
 \tilde{\Psi}^{(0)} (\lambda,\rho,z) = \mbox{diag} && \left( \frac{(\mu_5-\lambda)}{(\mu_6-\lambda)} , \frac{(\mu_1-\lambda) (\bar{\mu}_2-\lambda) (\mu_3-\lambda) (\mu_6-\lambda)}{(\mu_4-\lambda) (\mu_5-\lambda) (\mu_7-\lambda)}  , \right. \nonumber \\ 
 &&\left.-\frac{(\mu_2-\lambda) (\mu_7-\lambda)}{(\mu_3-\lambda)} \right).
\end{eqnarray}
\item
Perform a 2-soliton transformation with $\tilde{g}^{(0)}$ as a seed: 
Add an anti-soliton at $z=a_1$ with BZ vector $m_0^{(1)} = (1,0,c_1)$ and
a soliton at $z=a_4$ with BZ vector $m_0^{(2)} = (1,0,c_2)$.
Denote the resulting metric $\tilde{g}$.
\item
Rescale $\tilde{g}$ to find
\begin{equation}
 g = - \frac{\mu_4}{\mu_1} \tilde{g}.
\end{equation}
\item
The conformal factor $f$ is constructed using Eq. (\ref{4-f-new}),
\begin{equation}
  f=f^{(0)} \frac{\det \Gamma}{\det \Gamma^{(0)}}.
\end{equation}
\end{enumerate}
The result $(g,f)$ is the final solution after performing an appropriate linear coordinate transformation. 
The exact expressions of the final metric are written down in the next subsection \ref{sec:solution1}.
The rod structure of regular black di-ring with single spin is given in Fig. \ref{fig:rod_diring}.

\subsubsection{Solution of black di-ring}
\label{sec:solution1}
We write down the metric functions of black di-ring.
The non-zero components of the metric can be written as
\begin{equation}
g_{tt} = \frac{X_1 +c_1{}^2 X_2 +c_2{}^2 X_3 + c_1 c_2 X_4 +c_1{}^2 c_2{}^2 X_5}{\mu _6\left(D_1 +c_1{}^2 D_2 +c_2{}^2 D_3 + c_1 c_2 D_4 +c_1{}^2 b_c{}^2 D_5\right)},
\end{equation}
\begin{equation}
g_{t\phi} = \frac{c_1 Y_1 +c_2 Y_2 + c_1{}^2c_2 Y_3 +c_1 c_2{}^2 Y_4}{\left(D_1 +c_1{}^2 D_2 +c_2{}^2 D_3 + c_1c_2 D_4 +c_1{}^2 c_2{}^2 D_5\right)},
\end{equation}
\begin{equation}
g_{\phi\phi} = \frac{Z_1 +c_1{}^2 Z_2 +c_2{}^2 Z_3 + c_1c_2 Z_4 +c_1{}^2 c_2{}^2 Z_5}{\mu _1\mu _3\left(D_1 +c_1{}^2 D_2 +c_2{}^2 D_3+ c_1c_2 D_4 +c_1{}^2 c_2{}^2 D_5\right)},
\end{equation}
\begin{equation}
g_{\psi\psi} = \frac{\rho ^2\mu _3\mu _6}{\mu _2\mu _5\mu _7},
\end{equation}
and
\begin{equation}
f=\frac{c_2{}^2}{\left(c_1-c_2\right){}^2} \frac{D_1+c_1{}^2 D_2 +c_2{}^2 D_3 + c_1c_2 D_4 +c_1{}^2 c_2{}^2 D_5}{\mu _3^2 \mu_4 \mu _5^2 M_{24}^2 M_{46}^2 M_{47}^2 W_{14}{}^2 W_{12}{}^2 W_{16}{}^2 W_{17}{}^2} f^{(0)}.
\end{equation}
The definitions of function $D_i$'s are 
\begin{equation}
D_1=\mu _3^2  \mu _4  \mu _5^2 M_{24}{}^2 M_{46}{}^2  M_{47}{}^2 W_{12}{}^2 W_{14}{}^2 W_{16}{}^2 W_{17}{}^2,
\end{equation}
\begin{equation}
D_2=\rho ^2 \mu _1^2 \mu _2 \mu _3 \mu _4 \mu _5 \mu _6 \mu _7  M_{14}{}^2 M_{24}{}^2  M_{46}{}^2 M_{47}{}^2 W_{13}{}^2 W_{15}{}^2 ,
\end{equation}
\begin{equation}
D_3=\rho ^2 \mu _2 \mu _3 \mu _4  \mu _5 \mu _6 \mu _7 M_{14}{}^2 M_{34}{}^2 M_{45}{}^2 W_{12}{}^2 W_{16}{}^2 
W_{17}{}^2,
\end{equation}
\begin{equation}
D_4=2 \mu _1 \mu _2 \mu _3 \mu _4 \mu _5 \mu _6  \mu _7  M_{24} M_{34}  M_{45}  M_{46}  M_{47} W_{11} W_{12} W_{13} W_{44} W_{15} W_{16}  W_{17},
\end{equation}
and
\begin{equation}
D_5=\mu _1^2 \mu _2^2  \mu _4 \mu _6^2 \mu _7^2 M_{34}{}^2 M_{45}{}^2  W_{13}{}^2 W_{14}{}^2  W_{15}{}^2,
\end{equation}
where
\begin{equation}
 M_{pq} = \mu_p-\mu_q.
\end{equation}
The functions $X_i$'s  are defined by the  function $D_i$'s as
\begin{equation}
X_1=-\frac{\mu _1 \mu _5}{\mu _4}D_1,
\end{equation}
\begin{equation}
X_2=\frac{\rho ^2 \mu _5}{\mu _1 \mu _4}D_2,
\end{equation}
\begin{equation}
X_3=\frac{\mu _1 \mu _4 \mu _5}{\rho ^2}D_3,
\end{equation}
\begin{equation}
X_4=-\mu _5 D_4,
\end{equation}
and
\begin{equation}
X_5=-\frac{\mu _4 \mu _5}{\mu _1}D_5.
\end{equation}
The functions $Y_i$'s are defined as
\begin{equation}
Y_1=-\mu _2 \mu _3 \mu _4  \mu _5^2  \mu _7 M_{14} M_{24}{}^2  M_{46}{}^2 M_{47}{}^2  W_{11}  W_{12} W_{13}  W_{14}W_{15} W_{16} W_{17} ,
\end{equation}
\begin{equation}
Y_2=\mu _2  \mu _3  \mu _5^2  \mu _7 M_{14} M_{24} M_{34}  M_{45} M_{46} M_{47} W_{12}{}^2 W_{14} W_{44}  W_{16}{}^2 W_{17}{}^2,
\end{equation}
\begin{equation}
Y_3=-\mu _1^2 \mu _2^2 \mu _5 \mu _6 \mu _7^2 M_{14} M_{34} M_{24}  M_{45} M_{46} M_{47} W_{13}{}^2 W_{14} W_{44}  W_{15}{}^2 ,
\end{equation}
and
\begin{equation}
Y_4=  \mu _2^2 \mu _4 \mu _5 \mu _6 \mu _7^2  M_{14} M_{34}{}^2 M_{45}{}^2 W_{11} W_{12} W_{13} W_{14}  W_{15}  W_{16} W_{17} .
\end{equation}
The functions $Z_i$'s  are defined by the  function $D_i$'s as
\begin{equation}
Z_1=\mu _2 \mu _4 \mu _7 D_1,
\end{equation}
\begin{equation}
Z_2=-\frac{\mu _1^2 \mu _2 \mu _4 \mu _7}{\rho ^2} D_2,
\end{equation}
\begin{equation}
Z_3=-\frac{\rho ^2 \mu _2 \mu _7}{\mu _4} D_3,
\end{equation}
\begin{equation}
Z_4=\mu _1 \mu _2 \mu _7 D_4,
\end{equation}
and
\begin{equation}
Z_5=\frac{\mu _1^2 \mu _2 \mu _7}{\mu _4}D_5.
\end{equation}


%

To remove singularities at $z=a_1$ and $z=a_4$, we set the BZ parameters 
\begin{equation}
c_1= \pm \sqrt{\frac{2 a_{21} a_{61} a_{71}}{a_{31} a_{51}}},
\end{equation}
\begin{equation}
c_2= \pm \sqrt{\frac{2 a_{42} a_{64} a_{74}}{a_{43} a_{54}}},
\end{equation}
where $a_{pq}=a_p-a_q$.

Note that, in order for the metric to asymptotically approach the Minkowski spacetime without global rotation, we have to replace $\phi\phi$ and $t\phi$ components of the metric by using the freedom of linear coordinate transformation as,
\begin{equation}
 g_{\phi\phi} \rightarrow g_{\phi\phi} + 2 C g_{t\phi} + C^2 g_{tt}, ~~~ g_{t\phi} \rightarrow g_{t\phi} + C g_{tt}
\end{equation}
where
\begin{equation}
 C = \frac{2 a_{41} a_{43} a_{45} c_2}{2 a_{42} a_{64} a_{74} - a_{43} a_{54} c_1 c_2}.
\end{equation}
Using the exact expressions of black di-ring, we can compute physical variables of black di-ring.
For example,  the angular velocities of outer and inner horizons which appear in the direction vectors of rod structure 
of Fig. \ref{fig:rod_diring} are obtained as 
\begin{equation}
\Omega_1=\frac{(a_{42}c_1+a_{21}c_2)(c_1-c_2)}{a_{41}(2(a_{42}c_1+a_{21}c_2)-c_1 c_2 (c_1-c_2))},
\end{equation}
and
\begin{equation}
\Omega_2=\frac{a_{41}c_1 c_2 (c_1-c_2)}{2\left((a_{64}a_{74}c_1-a_{61}a_{71}c_2)(c_1-c_2)+a_{41}^2 c_1 c_2\right)}.
\end{equation}

\subsubsection{Physical properties of  black di-ring}
In this subsection we briefly introduce the results of investigations for the physical properties of black di-ring.
As similar as the black Saturn, one of the most important feature of black di-ring is continuous non-uniqueness.
This fact can be confirmed by a plot of random systematical sampling in the phase diagram.

Previously, it was argued that it is unlikely that multi-ring solutions exist as states in thermodynamical equilibrium because a black ring is uniquely determined by fixing the temperature $T$ and angular velocity $\Omega$. 
Contrary to this expectation, it was shown that the thermodynamical equilibrium black di-ring is possible.\cite{Iguchi:2010pe} 
Requiring equal temperatures and angular velocities for the both horizons imposes two conditions on the parameters. 
As a result, thermodynamical equilibrium black di-ring form a curve in the phase diagram.

The black di-ring in thermodynamic equilibrium is not thermodynamically stable because the entropy of it is not the maximum. 
In addition, a possibility that the thermodynamic equilibrium is local maximum of the entropy is denied 
by the numeric search.\cite{Iguchi:2010pe}
While the window of the metastability is closed, it was found that there is a narrow window $0.92075 < j^2 < 0.92084$ of the thin ring branch where the both eigenvalues 
of Hessian matrix of the entropy function are positive. 
In this region the entropy of the thermodynamic equilibria is locally minimum.  
It is likely that the interaction between the black rings makes the system thermodynamically unstable.

\subsection{Orthogonal black di-ring solution}

Here, we construct the solution with two black rings whose planes of 
$S^1$ rotations are orthogonal to each other. 
In the five-dimensional spacetime, we can take two spacelike orthogonal planes. 
In the orthogonal black di-ring solution, on each plane a $S^1$ rotating black ring exists. 

\begin{figure}[t]
  \begin{center}
    \includegraphics[keepaspectratio=true,height=30mm]{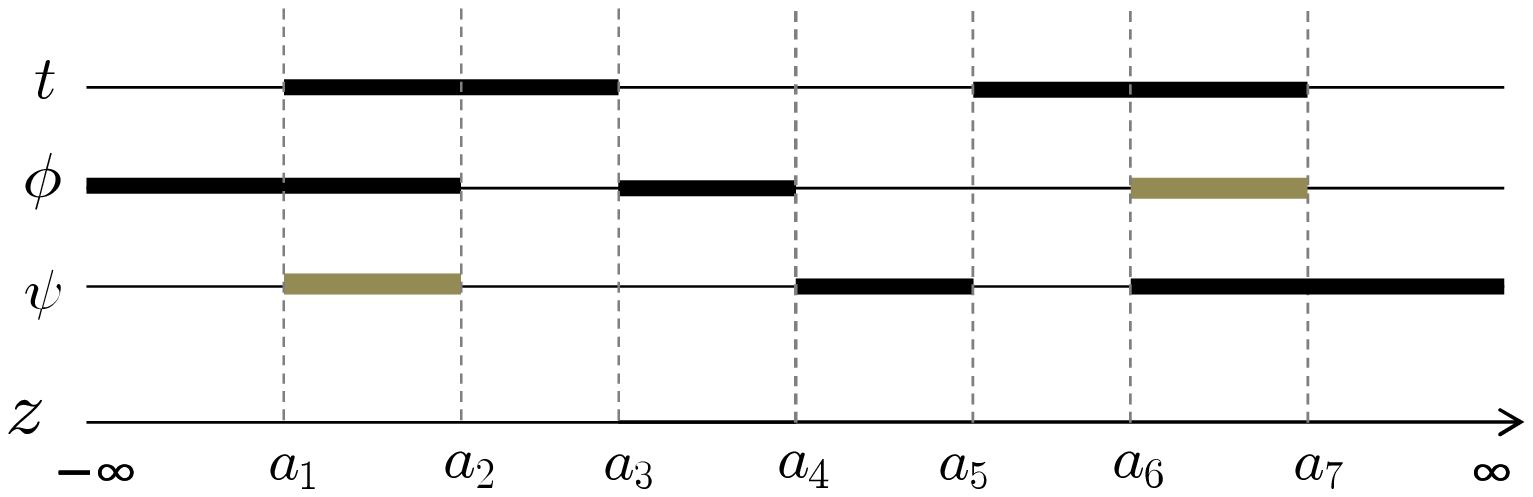}
  \end{center}
  \caption
{Rod structure of seed metric of orthogonal black di-ring: 
rods with  linear mass density $1/2$ are described as the thick 
black lines. The thick gray line means  linear mass density  $-1/2$. }
  \label{4-figorthogonal1}
\end{figure}

The orthogonal black di-ring solution was constructed with the seed solution
\cite{Izumi:2007qx}
\begin{eqnarray}
g_{ij}^{(0)}=\mbox{diag} \left( 
-\frac{\mu_1 \mu_5}{\mu_3\mu_7} ,
\frac{\rho^2 \mu_3 \mu_7}{ \mu_2 \mu_4 \mu_6},
\frac{\mu_2\mu_4\mu_6}{\mu_1\mu_5}
\right)
\end{eqnarray}
and
\begin{eqnarray}
&&f^{(0)}= k^2\frac{\mu_2\mu_4\mu_6}{\mu_1\mu_5}
\frac{W_{12}W_{13}W_{14}W_{16}W_{17}W_{23}W_{25}W_{27}W_{34}W_{35}W_{36}W_{45}W_{47}
W_{56}W_{57}W_{67}}{W^{\ \ 2}_{24}W^{\ \ 2}_{26}W^{\ \ 2}_{37}W^{\ \ 2}_{46}\prod_{p=1}^7W_{pp}},\nonumber\\
&&{}
\end{eqnarray}
whose rod structure is described as Fig. \ref{4-figorthogonal1}.
Removing a soliton $\mu_7$ and an anti-soliton ${\bar\mu}_1$ with trivial BZ vectors 
and rescaling it by a factor $\mu_1/\mu_7$,
we have 
\begin{eqnarray}
{\tilde g}^{(0)} =
 \mbox{diag} \left( 
-\frac{\mu_5}{\mu_3} ,
\frac{\rho^2 \mu_1 \mu_3}{ \mu_2 \mu_4 \mu_6},
\frac{\mu_2\mu_4\mu_6}{\mu_5\mu_7}
\right).\label{4-tildeg}
\end{eqnarray}
and the generating matrix ${\tilde \Psi}^{(0)}$ is written as 
\begin{eqnarray}
&&
{\tilde \Psi}^{(0)} =
 \mbox{diag}\biggl(
-\frac{(\mu_5-\lambda)}{(\mu_3-\lambda)} ,
-\frac{ (\mu_1-\lambda) (\mu_3-\lambda)(\bar\mu_4-\lambda)}
{ (\mu_2-\lambda) (\mu_6-\lambda)},
\nonumber\\
&&\qquad\qquad\qquad\qquad\qquad\qquad\qquad
\frac{(\mu_2-\lambda)(\mu_6-\lambda)(\bar\mu_7-\lambda)}
{(\mu_5-\lambda)(\bar\mu_4-\lambda)}
\biggr).
\end{eqnarray}
Then, we perform a 2-soliton transformation with $\tilde g^{(0)}$ as seed metric. 
Here, we use a soliton $\mu_7$ with a BZ vector $(1,b,0)$ 
and an anti-soliton ${\bar\mu}_1$ with a BZ vector $(1,0,c)$.
Rescaling the obtained metric by a factor $\mu_7/\mu_1$, we have the orthogonal 
black di-ring solution.   
The scalar function $f$ is obtained as
\begin{eqnarray}
f=f^{(0)} \frac{\det\Gamma}{\det\Gamma^{(0)}},
\end{eqnarray}
where $\Gamma^{(0)}=\Gamma |_{b=0,c=0}$.

The metric of the orthogonal black di-ring solution is written as 
\begin{eqnarray}
&&f= f_0 H/F,\\
&&H=F+b^2F_{(b)}+c^2F_{(c)}+b^2c^2F_{(bc)},\\
&&F=-\frac{\mu_5^2 M_{37}^{\ \ 2}W_{13}^{\ \ 2}W_{17}^{\ \ 2}}
{\rho^4 \mu_3^2 M_{57}^{\ \ 2}W_{15}^{\ \ 2} W_{11}W_{77}},\\
&&F_{(b)}=-\frac{\mu_1\mu_4\mu_5M_{27}^{\ \ 2}M_{67}^{\ \ 2}W_{13}^{\ \ 2}}
{\mu_2\mu_6 M_{37}^{\ \ 2}W_{15}^{\ \ 2}W_{47}^{\ \ 2}W_{11}W_{77}},\\
&&F_{(c)}=-\frac{\mu_1^2\mu_2\mu_6\mu_7M_{14}^{\ \ 2}M_{37}^{\ \ 2}W_{15}^{\ \ 2}}
{\rho^2 \mu_3\mu_4M_{57}^{\ \ 2}W_{12}^{\ \ 2}W_{16}^{\ \ 2}W_{11}W_{77}},\\
&&F_{(bc)}=\frac{\mu_1^3\mu_3\mu_7 M_{14}^{\ \ 2}M_{27}^{\ \ 2}M_{67}^{\ \ 2}W_{15}^{\ \ 2}}
{\mu_5 M_{17}^{\ \ 2}M_{37}^{\ \ 2}W_{12}^{\ \ 2}W_{16}^{\ \ 2}W_{47}^{\ \ 2}W_{11}W_{77}},
\end{eqnarray}
\begin{eqnarray}
&&g_{tt}=H^{-1}(A+b^2A_{(b)}+c^2A_{(c)}+b^2c^2A_{(bc)}),\\
&&A=\frac{\mu_1\mu_5^3 M_{37}^{\ \ 2}W_{13}^{\ \ 2}W_{17}^{\ \ 2}}
{\rho^4\mu_3^3\mu_7 M_{57}^{\ \ 2}W_{15}^{\ \ 2}W_{11}W_{77}},\\
&&A_{(b)}=-\frac{\mu_1^2\mu_4\mu_5^2\mu_7 M^{\ \ 2}_{27}M^{\ \ 2}_{67}W^{\ \ 2}_{13}}
{\rho^2\mu_2\mu_3\mu_6 M^{\ \ 2}_{37}W^{\ \ 2}_{15}W^{\ \ 2}_{47}W_{11}W_{77}},\\
&&A_{(c)}=-\frac{\mu_1\mu_2\mu_5\mu_6 M^{\ \ 2}_{14}M^{\ \ 2}_{37}W^{\ \ 2}_{15}}
{\mu_3^2\mu_4 M^{\ \ 2}_{57}W^{\ \ 2}_{12}W^{\ \ 2}_{16}W_{11}W_{77}},\\
&&A_{(bc)}=-\frac{\mu_1^2\mu_7^2 M^{\ \ 2}_{14}M^{\ \ 2}_{27}M^{\ \ 2}_{67}W^{\ \ 2}_{15}}
{M^{\ \ 2}_{17}M^{\ \ 2}_{37}W^{\ \ 2}_{12}W^{\ \ 2}_{16}W^{\ \ 2}_{47}W_{11}W_{77}},
\end{eqnarray}
\begin{eqnarray}
&&g_{\phi\phi}= B+b^2 H^{-1}( B_{(b)}  + c^2 B_{(bc)}),\\
&&B=\frac{\rho^2\mu_3\mu_7}{\mu_2\mu_4\mu_6},\\
&&B_{(b)}=\frac{\rho^2\mu_1\mu_3\mu_5M^{\ \ 2}_{27}M^{\ \ 2}_{67}W^{\ \ 2}_{13}}
{\mu_2^2\mu_6^2\mu_7 M^{\ \ 2}_{37}W^{\ \ 2}_{15}W^{\ \ 2}_{47}W_{11}},\\
&&B_{(bc)}=-\frac{\rho^2\mu_1^3\mu_3^2M^{\ \ 2}_{14}M^{\ \ 2}_{27}M^{\ \ 2}_{67}W^{\ \ 2}_{15}}
{\mu_2\mu_4\mu_5\mu_6M^{\ \ 2}_{17}M^{\ \ 2}_{37}W^{\ \ 2}_{12}W^{\ \ 2}_{16}W^{\ \ 2}_{47}W_{11}},
\end{eqnarray}
\begin{eqnarray}
&&g_{\psi\psi}=C+c^2 H^{-1}( C_{(c)}  +b^2  C_{(bc)}),\\
&&C=\frac{\mu_2 \mu_4 \mu_6}{\mu_1 \mu_5},\\
&&C_{(c)}=\frac{\mu_1\mu_2^2\mu_6^2\mu_7M^{\ \ 2}_{14}M^{\ \ 2}_{37}W^{\ \ 2}_{15}}
{\rho^4\mu_3\mu_5M^{\ \ 2}_{57}W^{\ \ 2}_{12}W^{\ \ 2}_{16}W_{77}},\\
&&C_{(bc)}=-\frac{\mu_1^2\mu_2\mu_3\mu_4\mu_6\mu_7 M^{\ \ 2}_{14}M^{\ \ 2}_{27}M^{\ \ 2}_{67}W^{\ \ 2}_{15}}
{\rho^2 \mu_5^2 M^{\ \ 2}_{17}M^{\ \ 2}_{37}W^{\ \ 2}_{12}W^{\ \ 2}_{16}W^{\ \ 2}_{47}W_{77}},
\end{eqnarray}
\begin{eqnarray}
&&g_{t\phi} = H^{-1} b(D +c^2 D_{(c)}),\\
&&D=\frac{\mu_1\mu_5^2M_{27}M_{67}W^{\ \ 2}_{13}W_{17}}
{\rho^2\mu_2\mu_3\mu_6\mu_7 M_{57}W^{\ \ 2}_{15}W_{47}W_{11}},\\
&&D_{(c)}=-\frac{\mu_1^2 M^{\ \ 2}_{14}M_{27}M_{67} W^{\ \ 2}_{15}}
{\mu_4 M_{17}M_{57}W^{\ \ 2}_{12}W^{\ \ 2}_{16}W_{47}W_{11}},
\end{eqnarray}
\begin{eqnarray}
&&g_{t\psi} = H^{-1} c(E +b^2 E_{(b)}),\\
&&E=\frac{\mu_2\mu_5\mu_6 M_{14}M^{\ \ 2}_{37}W_{13}W_{17}}
{\rho^4\mu_3^2M^{\ \ 2}_{57}W_{12}W_{16}W_{77}},\\
&&E_{(b)}=-\frac{\mu_1\mu_4\mu_7 M_{14} M^{\ \ 2}_{27}M^{\ \ 2}_{67}W_{13}}
{\rho^2M_{17}M^{\ \ 2}_{37}W_{12}W_{16}W^{\ \ 2}_{47}W_{77}}, 
\end{eqnarray}
\begin{eqnarray}
g_{\phi \psi}=H^{-1}bc \frac{\mu_1M_{14}M_{27}M_{67}W_{13}}
{\rho^2 M_{17}M_{57}W_{12}W_{16}W_{47}}.
\end{eqnarray}

This solution has generally singularities at 
$(\rho,z)=(0,a_1)$ and $(0,a_7)$
and conical singularities on the rods $z \in[-\infty,a_2]$,
$z \in[a_3,a_4]$, $z \in[a_4,a_5]$ and $z \in[a_6,\infty]$.
These singularities can be removed if we set the BZ parameters 
\begin{eqnarray}
&&b=\pm\sqrt{\frac{2a_{71}a_{73}^{\ \ 2}a_{74}}{a_{72}a_{75}a_{76}}},\\
&&c=\pm\sqrt{\frac{2a_{21}a_{31}a_{61}a_{71}}{a_{41}a_{51}^{\ \ 2}}},
\end{eqnarray}
and the integration constant in $f$ 
\begin{eqnarray}
 k^2=1,
\end{eqnarray}
and if we impose the balance conditions 
\begin{eqnarray}
&&1 = \frac{
a_{14}\, a_{16}\, a_{17}\, a_{25}\, a_{27}\, a_{34}\, a_{35}\, a_{36}}
{a_{15}^{\ 2}\, a_{24}^{\ 2}\, a_{26}^{\ 2}\, a_{37}^{\ 2}},\\
&&1 = \frac{a_{74}\, a_{72}\, a_{71}\, a_{63}\, a_{61}\, a_{54}\, a_{53}\, a_{52}}
{a_{73}^{\ 2}\, a_{64}^{\ 2}\, a_{62}^{\ 2}\, a_{51}^{\ 2}}.
\end{eqnarray}

\begin{figure}[t]
  \begin{center}
    \includegraphics[keepaspectratio=true,height=30mm]{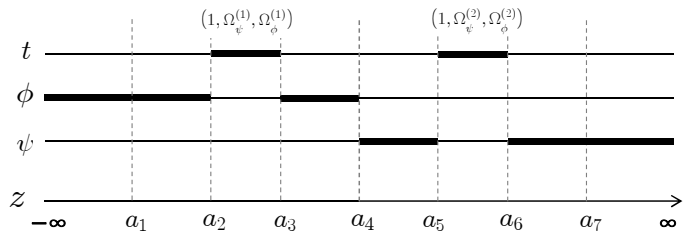}
  \end{center}
  \caption
{Rod structure of regular orthogonal black di-ring: 
rods with  linear mass density $1/2$ are described as the thick 
black lines. }
  \label{4-figorthogonal2}
\end{figure}

The rod structure of the obtained regular metric 
is illustrated in Fig. \ref{4-figorthogonal2}.
The directions of rods on horizons are 
\begin{eqnarray}
&&\left( 1,\Omega_\psi^{(1)},\Omega_\phi^{(1)}\right) \qquad \mbox{on } z\in [a_2,a_3],\\ 
&&\left(1,\Omega_\psi^{(2)},\Omega_\phi^{(2)}\right)\qquad \mbox{on } z\in [a_4,a_5], 
\end{eqnarray}
where
\begin{eqnarray}
&&\Omega_\psi^{(1)}=-\frac{b\, a_{75}\, a_{76}}{2\, a_{73}^{\ \ 2}\, a_{71}},\qquad
\Omega_\phi^{(1)}=-\frac{c\, a_{51}^{\ \ 2}}{2\, a_{31}\, a_{61}\, a_{71}},
\\
&&\Omega_\psi^{(2)}=-\frac{b\, a_{76}}{2\, a_{71}\, a_{74}}, \qquad
\Omega_\phi^{(2)}=-\frac{c\, a_{41}}{2\, a_{61}\, a_{71}}.
\end{eqnarray}
This regular solution has four parameters which are corresponding to 
the radii and the angular momenta of the $S^1$ rotations of both black rings. 
Each black ring rotates to the $S^2$ direction because of the drag by 
the $S^1$ rotation of the other black ring. 

\subsection{Black multi-ring}

It is expected that we can construct the solution where 
there are superposed multiple $S^1$ rotating black rings. 
Here, we show the expected method to construct the solution. 

\begin{figure}[t]
  \begin{center}
    \includegraphics[keepaspectratio=true,height=60mm]{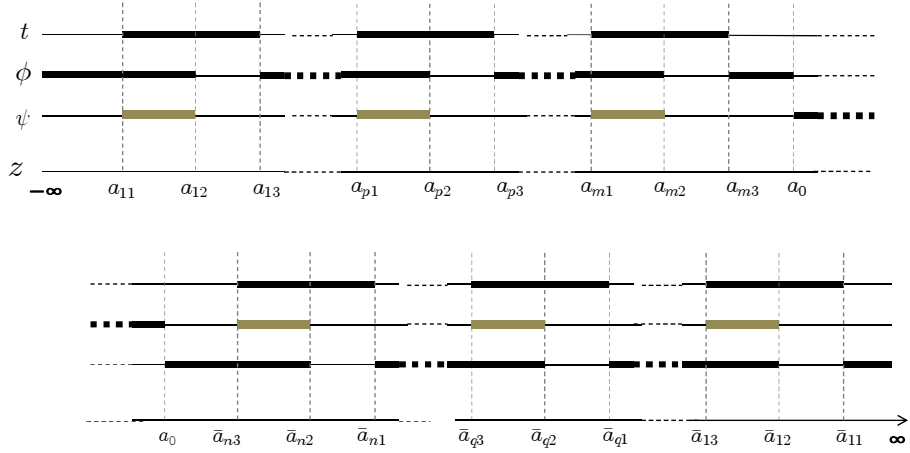}
  \end{center}
  \caption
{Rod structure of seed metric of black multi-ring: 
rods with  linear mass density $1/2$ are described as the thick 
black lines. The thick gray line means  linear mass density  $-1/2$. 
 }
  \label{4-multiseed}  
\end{figure}

We try to construct the solution where $m+n$ black rings exist. 
We consider the situation where
$S^1$ planes of $m$ black rings are the same and orthogonal to 
the $S^1$ planes of the other $n$ black rings. 
 From the construction of $S^1$ rotating solutions in the previous subsections, 
it is suspected that we should introduce the non-trivial structure 
outside of each black ring in the seed metric. 
Then, the seed metric is described as Fig. \ref{4-multiseed} and 
written as 
\begin{eqnarray}
g^{(0)}_{ij}=\mbox{diag} \left(
-\prod_{p=1}^m \frac{\mu_{p1}}{\mu_{p3}}
\prod_{q=1}^n \frac{\mu_{\bar q3}}{\mu_{\bar q1}},
\frac{\rho^2}{\mu_0}\prod_{p=1}^m \frac{\mu_{p3}}{\mu_{p2}}
\prod_{q=1}^n \frac{\mu_{\bar q2}}{\mu_{\bar q3}},
\mu_0\prod_{p=1}^m \frac{\mu_{p2}}{\mu_{p1}}
\prod_{q=1}^n \frac{\mu_{\bar q1}}{\mu_{\bar q2}}
\right),
\label{4-multi}
\end{eqnarray}
where $\mu_{\bar qi}$ is the soliton constructed with the constant $\bar a_{qi}$.

Starting from the seed metric Eq. (\ref{4-multi}), 
we remove $m$ anti-solitons $\bar \mu_{p1}$ ($p=1\cdots m$) and 
$n$ solitons $\mu_{\bar q 1}$ ($q=1\cdots n$). 
After that we perform the transformation with $m$ anti-solitons $\bar \mu_{p1}$ 
and  $n$ solitons $\mu_{\bar q 1}$ where 
BZ vectors associated with the anti-solitons $\bar \mu_{p1}$ and 
with the solitons $\mu_{\bar q 1}$are $(1,0,c_p)$ and $(1,b_q,0)$, respectively. 
Generally, the obtained solution has singularities at $z=a_{p1}$ and $z=\bar a_{q1}$ 
and conical singularities on the axes. 
It is expected that these singularities can be removed by tuning the parameters 
$b_q$, $c_p$, $a_{p1}$ and $\bar a_{q1}$ and that 
the regular solution has $2m+2n$ parameters. 
They are corresponding to the radii and the angular momenta of
the $S^1$ direction of $m+n$ black rings.

\section*{Acknowledgements}

H.I. is  supported by Grant-in-Aid for Young Scientists (B)
(No. 20740143) from Japanese Ministry of Education, Science,
Sports, and Culture.

K.I. acknowledges supports by the Grant-in-Aid for Scientific Research (A) No. 21244033
and Japan-Russia Research Cooperative Program.


%

\end{document}